\documentclass[traditabstract]{aa} % for the abstract without structuration (traditional abstract) \usepackage[toc,page]{appendix}usepackage{graphicx}
\usepackage{natbib}
\usepackage{multirow}
\bibpunct{(}{)}{;}{a}{,}{,}
\usepackage{txfonts}
\usepackage[colorlinks=true,     linkcolor=blue, citecolor=blue, filecolor=blue, urlcolor=blue]{hyperref}

\begin{document}
   \title{Gas phase Elemental abundances in Molecular cloudS (GEMS)}
   \subtitle{ IX. Deuterated compounds of H$_2$S in starless cores.}

   \author{
    M. Rodr{\'{\i}}guez-Baras\inst{1}
    \and
    G. Esplugues\inst{1}
    \and
    A.~Fuente\inst{2}
    \and
    S. Spezzano\inst{3}
    \and
    P. Caselli\inst{3}
    \and
    J.C. Loison\inst{4}
    \and
    E. Roueff\inst{5}
    \and
    D. Navarro-Almaida\inst{6}
    \and
    R. Bachiller\inst{1}
    \and
    R. Mart\'{i}n-Dom\'{e}nech\inst{2}
    \and
    I. Jim\'{e}nez-Serra\inst{2}
    \and
    L. Beitia-Antero\inst{7}
    \and
    R. Le Gal\inst{8}
    }
             
   \institute{Observatorio Astron\'omico Nacional (OAN), Alfonso XII, 3,  28014, Madrid, Spain
    \and
    Centro de Astrobiolog\'{\i}a (CSIC-INTA), Ctra. de Ajalvir, km 4, Torrej\'on de Ardoz, 28850, Madrid, Spain
    \and
    Centre for Astrochemical Studies, Max-Planck-Institute for Extraterrestrial Physics, Giessenbachstrasse 1, 85748, Garching, Germany
    \and
    Institut des Sciences Mol\'eculaires (ISM), CNRS, Univ. Bordeaux, 351 cours de la Lib\'eration, F-33400, Talence, France
    \and
    LERMA, Observatoire de PARIS, PSL Research University, CNRS, Sorbonne Universit\'e, 92190 Meudon, France
  	\and
    D\'epartement d'Astrophysique (DAp), Commissariat \`a l'\'Energie Atomique et aux \'Energies Alternatives (CEA), Orme des Merisiers, B\^at. 709, 91191 Gif sur Yvette, Paris-Saclay, France
    \and
    Departamento de Estad\'istica e Investigaci\'on Operativa, Facultad de Ciencias Matem\'aticas, Universidad Complutense de Madrid. Plaza de las Ciencias 3, 28040, Madrid, Spain  
    \and
    Harvard-Smithsonian Center for Astrophysics, 60 Garden St., Cambridge, MA 02138, USA
 }

 \abstract  {H$_2$S is predicted to form by hydrogenation of atomic sulphur on grains and it is thought to be the main sulphur reservoir in the ice, being therefore a key molecule to understand sulphur chemistry in the star formation process and to solve the missing sulphur problem in molecular clouds and star-forming regions. The study of H$_2$S deuterium fraction can be used to constrain this molecule formation pathways.} 
 {The aim of this work is to investigate for the first time the deuteration of H$_2$S in a large sample of starless cores. }
 {We use observations of the GEMS IRAM 30m Large Program and complementary IRAM 30m telescope observations. We consider a sample of 19 starless cores located in Taurus, Perseus, and Orion molecular clouds, detecting HDS in 10 of these starless cores, and D$_2$S in five. The H$_2$S single and double deuterium fractions are analysed with regard to their relation with the cloud physical parameters, their comparison with values obtained for other interstellar sources, and their comparison with deuterium fractions in early stage star-forming sources of abundant molecules: c-C$_3$H$_2$, H$_2$CS, H$_2$O, H$_2$CO, and CH$_3$OH. } 
 {We obtain a range of X(HDS)/X(H$_2$S)$\sim$0.025$-$0.2 in the starless cores with HDS detections. The five starless cores with D$_2$S detections show values of X(D$_2$S)/X(HDS)$\sim$0.05$-$0.3. H$_2$S single deuteration shows an inverse relation with the cloud kinetic temperature, but no trend is found with molecular hydrogen density or visual extinction. H$_2$S deuteration values in starless cores are similar to those observed in Class 0, although this may be a consequence of an observational bias due to the limited spatial resolution. Comparison with c-C$_3$H$_2$, H$_2$CS, H$_2$O, H$_2$CO, and CH$_3$OH in other interstellar sources reveals a general trend of decreasing deuteration with increasing temperature, with lower values for Class I and massive star-forming sources. In starless cores and Class 0 objects H$_2$CS and H$_2$CO present higher deuteration fractions than c-C$_3$H$_2$, H$_2$S, H$_2$O, and CH$_3$OH. H$_2$O shows single and double deuteration values one order of magnitude lower than those of H$_2$S and CH$_3$OH. } 
 {Differences between c-C$_3$H$_2$, H$_2$CS and H$_2$CO deuterium fractions and those of H$_2$S, H$_2$O, and CH$_3$OH are related to deuteration processes produced in gas or solid phases, respectively. We interpret the differences between H$_2$S and CH$_3$OH deuterations and that of H$_2$O as a consequence of differences on the formation routes in the solid phase, and can particularly be explained in terms of the different occurrence of the D-H and H-D substitution reactions in the ice, together with the chemical desorption processes. Further interferometric observations and laboratory experiments are needed to understand the deuteration processes.}
  
   \keywords{Astrochemistry -- ISM: abundances -- ISM: kinematics and dynamics -- ISM: molecules --
   stars: formation -- stars: low-mass}
   
 \maketitle  

%%%%%%%%%%%%%%%%%%%

\section{Introduction}
\label{Sec: Introduction}

Stars are mainly formed in giant molecular clouds and dark cloud complexes. The cloud filamentary structure is believed to funnel interstellar gas and dust into increasingly denser concentrations, which consequently contract and fragment, leading to gravitationally bound prestellar cores that eventually form stars. This star formation process is regulated by different physical phenomena, such as magnetic pressure, turbulence, and rotation. Gas chemistry plays a key role: the chemical composition of the gas regulates the temperature of the gas and its ionisation fraction. Molecular emission determines the cooling of the gas, which allows the fragmentation of molecular filaments into prestellar cores as a consequence of diminishing the thermal support relative to self-gravity. Furthermore, the gas chemical composition determines the ionisation fraction, which controls the coupling of magnetic fields with the gas, driving the dissipation of turbulence and angular momentum transfer, and therefore playing a crucial role in the cloud collapse and the dynamics of accretion disks \citep{Zhao2016, Padovani2013}. 

The chemical composition of the gas is, in turn, influenced by the star formation process itself. Therefore, its study in star forming regions is a powerful diagnostic tool to determine the different processes at work and their evolution with time. Particularly interesting is the study of the deuterium fractionation, that is the process that enriches the molecules in deuterium with respect to hydrogen \citep{Ceccarelli2014, Imai2018}. The estimated cosmic deuterium abundance relative to hydrogen is $\sim$ 1.5 $\times$ 10$^{-5}$ \citep{Linsky2003}, but its relative abundance in molecules is larger than this elemental D/H in very specific locations, and particularly in cold ($\sim$10 K) molecular clouds. At these temperatures the fastest reactions are those involving ions, as most neutral-neutral reactions have activation barriers and are generally slower. H$_{3}^{+}$ is one of the first formed molecular ions, as a product of the cosmic ray ionisation of H$_2$ and H. Deuterium fractionation is mainly caused by the reaction of H$_{3}^{+}$ with HD, the major reservoir of D-atoms, to form H$_2$D$^+$ and H$_2$ \citep[e.g.][]{Millar1989}. The formation of H$_2$D$^+$ proceeds without a reaction barrier, whereas the backward reaction with para-H$_2$ has an endothermicity of $\sim$232~K \citep{Watson1974, Caselli2012, Albertsson2013}.  In molecular clouds where the ortho-to-para H$_2$ ratio is $<$0.01, the H$_2$D$^+$/H$_{3}^{+}$ ratio becomes larger than the D/H elemental abundance for T$<$  20$-$30~K \citep[][]{Dislaire2012}. Similarly, D$_2$H$^{+}$ and D$_{3}^{+}$ are formed from H$_2$D$^+$ and D$_2$H$^{+}$, correspondingly, by reactions with HD. These ions react with other molecules in gas phase, possibly donating D$^{+}$ to neutral species, enhancing their D/H ratio. The enhanced atomic D/H ratio due to the dissociative recombination of the deuterated forms of H$_{3}^{+}$ is transferred to grains, where it deuterates the molecules on the surface. Maximum values of deuterium fraction are reached in dense cores (n(H$_2$)$>$10$^4$ cm$^{-3}$) with temperatures $\sim<$ 10$-$15~K, where CO molecules freeze on the grain surfaces. The molecular ion H$_3^+$ is efficiently destroyed by reactions with CO. When CO depletes from the gas-phase, the H$_{3}^{+}$ abundance is enhanced and the deuterium fraction is further increased \citep{Dalgarno1984, Caselli2002, Roberts2003}.

High values of deuterium fraction have  been found in moderate warm regions (30 $-$ 70~K) such as the Orion Bar where the H$_2$D$^+$/H$_3^+$ ratio is not expected to be enhanced relative to the cosmic D/H value \citep{Parise2009}. In these cases, gas-phase deuteration can occur via reactions with CH$_2$D$^+$ and C$_2$HD$^+$ \citep{Roueff2013, Roueff2015}. At these temperatures, HD may react with CH$^{+}_3$ and C$_2$H$^{+}_2$  whose reverse reactions have endothermicities higher than H$_2$D$^+$, leading to enhancements in the abundances of the deuterated ions  CH$_2$D$^+$ and C$_2$HD$^+$ \citep{Millar2005}. These ions react with some compounds such as HCN and H$_2$CO, which maintain high deuteration fractions in regions with temperatures of $\sim$50~K \citep{Parise2009}. 

The degree of deuterium fractionation is sensitive to temperature, CO depletion, H$_2$ density, and the ortho-to-para ratio of H$_2$. Therefore, it has been extensively employed as a good tool to study physical conditions and chemical states of molecular clouds \citep{Imai2018}. In general terms, the deuterium fraction is predicted to increase when the starless core evolves towards the onset of gravitational collapse, as the core density becomes more and more centrally peaked \citep[e.g.][]{Crapsi2007}. Afterwards, the deuterium fraction drops when the young stellar object formed at the core centre begins to heat its surroundings \citep[see e.g.][]{Emprechtinger2009, Ceccarelli2014}. 

The efficiency of the deuteration enrichment in cold core conditions is now well established with the detection of multiply deuterated molecules in starless cores and young protostars: ND$_2$H \citep{Roueff2000}, ND$_2$H and D$_2$CO \citep{Loinard2001}, ND$_2$ \citep{Melosso2020}, ND$_3$ \citep{Lis2002}, D$_2$S \citep{Vastel2003}, D$_2$CO \citep{Ceccarelli2002, Bacmann2003}, D$_2$CS \citep{Marcelino2005}, CHD$_2$OH and CD$_3$OH \citep{Parise2002, Parise2004}. Despite this wealth of detections, only a few sulphur-bearing deuterated molecules (HDCS, D$_2$CS, HDS and D$_2$S, and the molecular ion DCS$^{+}$) have been detected in dense cores and/or protostars \citep{vanDishoeck1995, Vastel2003, Marcelino2005, Fuente2016}, and in most cases the detections are restricted to a few prototypical objects. Different species probe different regions along the line of sight, and in particular S-bearing species are heavily depleted in the coldest and densest regions of starless cores \citep{Graedel1982, Agundez2013}, where N-bearing species such as NH$_3$, N$_2$H+ and nitriles (HCN, HNC, CN) are still abundant \citep{Tafalla2006, Sipila2018, Kim2020}. The observation of deuterated S-bearing species provides therefore different information on the physical conditions and chemical history of starless cores. 

The aim of this work is to investigate the deuteration of H$_2$S in cold cores. Since its first detection in W3, NGC 7538, and five other Galactic regions \citep{Thaddeus1972}, H$_2$S has been detected in the gas phase in several interstellar sources \citep[e.g.][]{Minh1989, vanDishoeck1995, Wakelam2004, Neufeld2015, RiviereMarichalar2021}. However, chemical models reveal that the observed interstellar abundance of H$_2$S cannot be explained by gas-phase synthesis alone. Models predict that, in the dense interstellar medium, atomic sulphur would stick on grains and be mostly hydrogenated to form H$_2$S \citep{Hatchell1998, Garrod2007, Esplugues2014, Oba2019}, which is thought to be the main sulphur reservoir in the ice \citep{Vidal2017, Navarro2020}. However, H$_2$S has never been detected in interstellar ices and its abundance relative to water is smaller than $\sim$1\% \citep{vanDishoeck1998, JimenezEscobar2011}. 
Deuterium fractionation can theoretically be used to constrain molecular formation pathways, because the ratios between the deuterated species and the main isotopologue are expected to be different if the molecule forms in the gas phase or as a result of grain surface chemistry. In the latter case, deuteration is expected to follow a statistical scheme \citep{Brown1989, Bacmann2020}. 

Gas phase Elemental abundances in Molecular CloudS (GEMS) is an IRAM 30m Large Program run between summer 2017 and 2019, which aims at measuring the gas-phase elemental abundances in a set of prototypical filaments \citep{Fuente2019, Navarro2020, RodriguezBaras2021, Spezzano2022a, Esplugues2022}. This work uses both GEMS and complementary IRAM 30m observations of a wide sample of cores located in several clouds, covering different star formation activity, allowing to explore the role of the environment on the dynamical and chemical evolution of the pre-stellar cores. This paper presents the analysis and comparison between H$_2$S, HDS, and D$_2$S observations, which provide essential information on the formation and deuteration of this species in starless cores. The paper is structured as follows: Section \ref{Sec: Observations} provides a summary of the observations. Section \ref{Sec: Observational sample} presents the observational sample of starless cores and the physical properties of the corresponding star-forming regions. The derivation of physical parameters of the starless cores is described in Sect. \ref{Sec: Physical parameters of the starless cores}. Section \ref{Sec: The H2S abundance} is focused on the analysis of the measured H$_2$S abundances. The emission of the H$_2$S deuterated compounds is analysed in Sect. \ref{Sec: Deuterated compounds: HDS and D2S}. The results obtained are discussed in Sect. \ref{Sec: Discussion}, and our summary and conclusions are given in Sect. \ref{Sec: Summary and conclusions}.

\section{Observations}
\label{Sec: Observations}

Part of the data employed in this work, particularly those corresponding to the emission of o-H$_2$S (transition 1$_{10}\rightarrow1_{01}$ at 168.763 GHz) and H$_2^{34}$S (transition 1$_{10}\rightarrow1_{01}$ at 167.911 GHz), were observed by the GEMS Large Program. These observations are thoroughly described in \citet{Fuente2019}, and only a summary is provided here. The 3 and 2 mm observations were carried out using the IRAM 30m telescope at Pico Veleta (Spain). The observing mode was frequency switching with a frequency throw of 6 MHz. The Eight MIxer Receivers (EMIR) and the Fast Fourier Transform Spectrometers (FTS) with a spectral resolution of 49 kHz ($\sim$0.09 km/s) were used.

This work also uses specific observations of the selected sources (described in Sect.~\ref{Sec: Observational sample}) focused on the emission of the H$_2$S deuterated compounds: HDS (transition 1$_{01}\rightarrow0_{00}$ at 244.555 GHz, with E$_u$ = 11.7 K) and o-D$_2$S (transition 1$_{11}\rightarrow0_{00}$ at 237.904 GHz, with E$_u$ = 11.4 K). These are the lowest lying transitions for the two molecules. These 1 mm observations were carried out with the IRAM 30m telescope between 8 and 14 September 2021, using the Eight MIxer Receivers (EMIR) and the Fast Fourier Transform Spectrometers (FTS) with a spectral resolution of 200 kHz ($\sim$0.24 km/s). The observing mode was frequency switching with a frequency throw of 12 MHz, adapted to remove standing waves between the secondary mirror and the receivers.  

The intensity scale is T$_{MB}$, which is related with T$_A^*$ by $T_{MB} =(F_{eff} / B_{eff})T_A^*$, where F$_{eff}$ is the telescope forward efficiency and B$_{eff}$ is the main beam efficiency\footnote{https://publicwiki.iram.es/Iram30mEfficiencies}. For the IRAM 30m telescope, the difference between T$_A^*$ and T$_{MB}$ scales is $\sim$27\% at 145 GHz and $\sim$56\% at 230 GHz. The half-power beam widths (HPBW) are 16 arcsec and 10.7 arcsec at 145 GHz and 230 GHz, respectively. The data reduction was carried out using GILDAS\footnote{https://www.iram.fr/IRAMFR/GILDAS/}/CLASS following a standard procedure. The line intensity errors, considering calibration, size source and other factors, are assumed to be $\sim$20\%, with the exception of two D$_2$S lines (sources B213-C1-1 and L1448-1), where the associated uncertainty is 25\%.

\section{Observational sample}
\label{Sec: Observational sample}

The observational sample of this work is composed of 19 starless cores that are a subset of those included in the sample of the GEMS program. The list of selected sources is shown in Table \ref{Table: Starless core sample}. The observed starless cores belong to different filaments of three nearby star forming regions: Taurus, Perseus and Orion. These molecular cloud complexes are characterised by having different star formation activity and therefore different illumination. The comparison of starless cores located in the same filament allows the investigation of the effect of time evolution on the chemistry of dark cores \citep[see e.g.][]{Frau2012}. The comparison of starless cores in different star forming regions provides information about the influence of UV radiation on the gas composition, and the general effect of the environment on the chemistry of the cores. The characteristics of the selected regions are thoroughly described in \citet{RodriguezBaras2021}. As a summary, the main properties of the considered filaments are the following: 

\textit{Taurus: B213/L1495.} The Taurus molecular cloud is located at a distance of 145 pc \citep{Yan2019}. With a total mass of 1.5$\times 10^4$ M$_{\odot}$ derived from CO data \citep{Pineda2010Taurus}, it is known to contain more than 250 young stellar objects. It is considered an archetype of exclusively low-mass star-forming regions. It has been the target of several cloud evolution and star formation studies \citep{Unge1987, Mizuno1995, Goldsmith2008}, being extensively mapped in CO \citep{Cernicharo1987, Onishi1996, Narayanan2008} and visual extinction \citep{Cam1999, Padoan2002, Schmalzl2010}. 

B213/L1495 is a Taurus prominent filament, a clear example of a star-forming region where the magnetic field lines are perpendicular to the main filament (Soler 2019). The morphology of the map with striations perpendicular to the filament suggests that the filament is accreting material from its surroundings \citep{Goldsmith2008, Palmeirim2013}. A population of dense cores are embedded in this filament \citep{Benson1989, Onishi2002, Tatematsu2004, Hacar2013, Punanova2018}. Some of these dense cores are starless, while others are associated with young stellar objects (YSOs) of different ages. Interestingly, the density of stars decreases from North to South suggesting a different dynamical/chemical age along the filament. GEMS observed 9 cuts along clumps \#1, \# 2,  \#5,  \#6,  \#7,  \#10,  \#12,  \#16, and  \#17 (core numbers from the catalogue of \citealp{Hacar2013}). All these cores except \#7 and \#17  were observed in the deuterated compounds of H$_2$S, and therefore included in this work. We use the H$_2$ column density and dust temperature maps of B213 obtained by \citet{Palmeirim2013} on the basis on the Herschel Gould Belt Survey \citep{Andre2010} and Planck data (c.f. \citealp{Bernard2010}) at an angular resolution of 18.2$"$. 

\textit{Perseus: Barnard 1, NGC 1333, IC348, L1448, and B5.} The Perseus molecular cloud is a well-known star-forming cloud, located at a distance ranging from 234 to 331~pc according to recent Gaia parallaxes and photometric data \citep{Zucker2020}. Perseus is the prototype of intermediate-mass star-forming region, where the infrared survey of \citet{Ladd1993} suggests intermediate properties between Taurus, with stars predominantly forming in relative isolation (YSO surface density $\sim$10 pc$^{-2}$), and the Orion complex, with large star-forming clusters with YSO surface densities of 100 pc$^{-2}$. The molecular cloud complex is associated with three clusters containing pre-main-sequence stars: IC 348, with an estimated age of 2~Myr \citep{Luhman2003}; NGC 1333, which is younger than 1Myr in age \citep{Lada1996, Wilking2004}; and the Per 0B2 association, which contains a B0.5 star \citep{Steenbrugge2003}. The molecular cloud itself contains numerous protostars and dense cores \citep{Hatchell2005, Hatchell2007b, Hatchell2007a}. GEMS observed 11 cuts along starless cores distributed in Barnard 1, IC348, L1448, NGC 1333 and B5. Those selected for this work are shown in Table \ref{Table: Starless core sample}. The group of cores in IC348 and NGC 1333 are close to the clusters and therefore immersed in a harsh environment, while Barnard 1 and L1448 are located in a quiescent region. The dust opacity and dust temperatures maps reported by \citet{Zari2016} are used in our analysis. In order to derive the molecular hydrogen column density from the dust opacity at 850$\mu$m ($\tau_{850}$), we used expression (7) of  \citet{Zari2016} and A$_V$=A$_K$/0.112. This expression give accurate values for low extinctions  (A$_V<$ 10 mag) but may underestimate their value towards the extinction peaks. In the range of values considered in Perseus, A$_V\sim$3$-$30 mag, the uncertainty in the values of A$_V$ is a factor of 2.

\textit{Orion A.} The Orion molecular cloud, at a distance of $\sim$428 pc \citep{Zucker2019}, is the most massive and most active star-forming complex in the local neighborhood. It is a large complex, composed itself by two giant molecular clouds: Orion A, hosting the HII region usually called the Orion nebula, and the more quiescent Orion B (see, e.g., \citealp{Pety2017}). Different clouds have been identified within Orion A based on millimeter, submillimeter, and infrared observations. The Orion molecular cloud 1 (OMC~1) was identified as dense gas directly associated with Orion KL \citep{Wilson1970, Zuckerman1973, Liszt1974}, then OMC~2 \citep{Gatley1974} and OMC~3 \citep{Kutner1976} were detected as subsequent clumps in CO emission located about 15$'$ and 25$'$ to the north of OMC~1. \citet{Bally1987} revealed that these clouds consist of the Integral-Shaped Filament (ISF) of molecular gas, part itself of a larger filamentary structure extending from North to South over 4$^\circ$. OMC~4 \citep{Johnstone1999} and OMC~5 \citep{Johnstone2006} are concentrations of submillimeter continuum emission in the southern part of the integral-shaped filament. GEMS observed three cuts along OMC-2 (ORI-C3), OMC-3 (ORI-C1), and OMC-4 (ORI-C2). These cuts avoid the protostars and stars in this active star forming region, probing different environments because of their different distance from the Orion nebula. In this work we consider the visual extinction peaks along cuts ORI-C1 and ORI-C2, while cut ORI-C3 was not observed in the deuterated compounds of H$_2$S. In our analysis, we use the dust opacity and dust temperatures maps reported by \citet{Lombardi2014}. The values of the molecular hydrogen column density are derived from the dust opacity at 850$\mu$m, using expression $A_{\rm K}=2640 \times \tau_{850} + 0.012$ of  \citet{Lombardi2014} and A$_V$=A$_K$/0.112.

\begin{table*}
\caption{Starless cores included in this work sample. Observation cuts associated to these starless cores in GEMS observations are indicated. \textbf{NOTES:} $^{(1)}$B213 core IDs are from \citet{Hacar2013}. IDs indicated in "Other names" column are from \citet{Onishi2002}. $^{(2)}$Perseus core IDs (L 1448, NGC 1333, Barnard 1, Barnard 5, IC348) are from \citet{Hatchell2007a}. $^{(3)}$NGC\,1333 core IDs indicated in "Other names" column are from \citet{Sandell2001}.}
\label{Table: Starless core sample}
\centering
\begin{tabular}{llrlll}\\
\hline\hline
\noalign{\smallskip}
Filament & Core & \multicolumn{2}{c}{Coordinates} & Other names & Cut \\
& ID & RA (J2000) & Dec (J2000) & & \\
\hline
\noalign{\smallskip}                          
\multicolumn{6}{c}{{\bf Taurus}} \\

B\,213$^{(1)}$  &  \#1  &  04:17:41.80   &     $+$28:08:47.0     & 5   &  C1 \\%   &   9 & 9  \\%&2.60$\cdot10^4$ \\%& 1.36$\cdot10^{12}$\\     
            &  \#2  &   04:17:50.60   &     $+$27:56:01.0     &  $-$ &  C2  \\% &   9 & 7  \\%&9.06$\cdot10^4$ \\%& 2.74$\cdot10^{11}$ \\   
            &  \#5  &  04:18:03.80   &     $+$28:23:06.0     & 7  &   C5 \\%  &  9 & 9  \\%&2.09$\cdot10^4$ \\%& 9.62$\cdot10^{11}$ \\    
            &  \#6  &  04:18:08.40   &     $+$28:05:12.0     &  8  &  C6 \\%  &   9 &  5  \\%& 8.96$\cdot10^4$ \\%& 7.80$\cdot10^{11}$\\    
%            &  \#7  &  04:18:11.50   &     $+$27:35:15.0     & 9  &  C7  \\%   &   9 & 8  \\%&1.03$\cdot10^4$ \\%& 1.38$\cdot10^{12}$\\    
            &  \#10 &  04:19:37.60   &     $+$27:15:31.0     & 13a &  C10  \\%  &   9 &  8  \\%&2.71$\cdot10^4$ \\%& 8.51$\cdot10^{11}$\\   
            &  \#12 &  04:19:51.70   &     $+$27:11:33.0     & $-$ &   C12  \\%  &  9 & 6  \\%&1.23$\cdot10^5$ \\%& 2.13$\cdot10^{11}$ \\  
            &  \#16 &  04:21:21.00   &     $+$27:00:09.0     & $-$ &  C16  \\%  &  9  &  9  \\%&4.77$\cdot10^4$ \\%& 7.88$\cdot10^{11}$ \\
%            &  \#17 &   04:27:54.00  &    $+$26:17:50.0      & 26b &   C17 \\%   &  9 & 5  \\%&8.64$\cdot10^4$ \\%& 3.26$\cdot10^{11}$ \\ 
\multicolumn{6}{c}{{\bf Perseus}} \\                   
L1448$^{(2)}$   &  \#32 &  03:25:49.00   &     $+$30:42:24.6      &   &    C1 \\%  & 8 & 7 \\%& 1.36$\cdot10^4$ \\%& 1.42$\cdot10^{12}$\\ 
%NGC\,1333$^{(2)}$   &       &   03:29:18.26   &      $+$31:28:02.0     &  &   C1 \\%   &   21 & 18 \\%& 6.73$\cdot10^4$ \\%& 2.77$\cdot10^{11}$\\    
%            &       &  03:28:41.60   &      $+$31:06:02.0     &   &   C2 \\% &  17 & 13 \\%& 7.31$\cdot10^4$ \\%& 5.60$\cdot10^{11}$ \\
NGC\,1333$^{(2)}$               &  \#46 &  03:29:11.00   &   $+$31:18:27.4     &   SK20$^{(3)}$  &  C3-1 \\% \multirow{2}{*}{C3} \\% & \multirow{2}{*}{17} & \multirow{2}{*}{11} \\%& 6.85$\cdot10^4$ \\%& 1.15$\cdot10^{12}$ \\ 
            &  \#60 & 03:28:39.40    &      $+$31:18:27.4     &   & C3-14  \\%   & & \\%& \\%&\\    
            &  \#51 &  03:29:08.80   &      $+$31:15:18.1     &   SK16    &    C4 \\% & 16  & 13  \\%& 5.74$\cdot10^4$ \\%& 1.81$\cdot10^{12}$ \\   
            &  \#53 & 03:29:04.50  &      $+$31:20:59.1     &   SK26   &   C5\\%  & 11 &  11  \\%& 1.44$\cdot10^5$ \\%& 5.41$\cdot10^{11}$\\   
            &  \#57 &  03:29:18.20     &      $+$31:25:10.8     &    SK33  &   C6 \\% & 9  & 9  \\%& 9.30$\cdot10^4$ \\%& 1.44$\cdot10^{11}$\\  
            &  \#64 &  03:29:25.50     &      $+$31:28:18.1     &  &   C7   \\%   & 1 & 1  \\%& 1.35$\cdot10^5$ \\%&  1.84$\cdot10^{11}$\\    
Barnard 1$^{(2)}$   &  1b   &  03:33:20.80 & $+$31:07:34.0  & &  C1 \\%  & 18  & 18 \\%& 6.20$\cdot10^4$ \\%& 1.95$\cdot10^{12}$  \\ 
%            &       &  03:33:01.90     &    $+$31:04:23.2    &  &    C2  \\%    & 8   & 6  \\%& 9.61$\cdot10^4$  \\%& 8.91$\cdot10^{11}$ \\
Barnard 5$^{(2)}$   &  \#79 &  03:47:38.99    &      $+$32:52:15.0    &   &    C1  \\%   &  13 &  9 \\%& 2.13$\cdot10^4$ \\%& 1.16$\cdot10^{12}$ \\    
IC\,348$^{(2)}$     &  \#1  &    03:44:01.00    &   $+$32:01:54.8   &  & C1-1\\% \multirow{2}{*}{C1} \\% & \textbf{\multirow{2}{*}{14}} & \textbf{\multirow{2}{*}{11}} \\%& \\% & \\    
%            &  \#10 &  03:44:05.74   &       $+$32:01:53.5      &    & C1-10 \\%   &   &  \\%& 6.83$\cdot10^4$  \\%& 2.48$\cdot10^{11}$\\ 
\multicolumn{6}{c}{{\bf Orion A}} \\                    
OMC-3     &       &   05:35:19.54   &      $-$05:00:41.5        & &   C1   \\%      & 20 & 10  \\%& 8.04$\cdot10^4$ \\%& 1.49$\cdot10^{12}$  \\  
OMC-4          &       &  05:35:08.15    &      $-$05:35:41.5        & &   C2    \\%     & 20 & 14  \\%& 4.48$\cdot10^4$ \\%& 5.98$\cdot10^{11}$ \\    
%            &       & 05:35:23.68  &     $-$05:12:31.8     &  &    C3    \\%     & 13 & 9  \\%& 1.58$\cdot10^5$ \\%& 7.75$\cdot10^{11}$ \\ 

\hline
\noalign{\smallskip}
%TOTAL       &              &    & &  &  & 305 & 244 \\                  
%\hline \hline
\end{tabular}

\end{table*}

\begin{table*}
\caption {Line parameters from Gaussian fits. NOTE: In the case of HDS and D$_2$S non-detections, the upper limit of the line intensity is estimated as three times the rms in temperature. The line area upper limit is obtained by multiplying the intensity upper limit by the width of the H$_2$S line measured in that spectrum. }
\label{Table: Line parameters}
\centering     
\resizebox{\textwidth}{!}{%    
\begin{tabular}{l|ccc|ccc|ccc|ccc}\\
\hline\hline
\noalign{\smallskip}
\multicolumn{1}{c}{} & \multicolumn{3}{c}{{H$_2$S 1$_{10}\rightarrow1_{01}$}}	&		\multicolumn{3}{c}{{H$_2^{34}$S 1$_{10}\rightarrow1_{01}$}}	&		\multicolumn{3}{c}{{HDS 1$_{01}\rightarrow0_{00}$}}	&	\multicolumn{3}{c}{{D$_2$S 1$_{11}\rightarrow0_{00}$}}\\
Starless core	&	V$_{\mathrm {LSR}}$	&	$\bigtriangleup$V	&	$T_{\mathrm {MB}}$	&	V$_{\mathrm {LSR}}$	&	$\bigtriangleup$V	&	$T_{\mathrm {MB}}$	&	V$_{\mathrm {LSR}}$	&	$\bigtriangleup$V	&	$T_{\mathrm {MB}}$ &		V$_{\mathrm {LSR}}$	&	$\bigtriangleup$V	&	$T_{\mathrm {MB}}$\\
	&	(km s$^{-1}$)	&	(km s$^{-1}$)	&	(K)	&	(km s$^{-1}$)	&	(km s$^{-1}$)	& (K) & (km s$^{-1}$)	&	(km s$^{-1}$)	&	(K) &	(km s$^{-1}$)	&	(km s$^{-1}$)	&	(K)\\
\noalign{\smallskip}
\hline
\noalign{\smallskip}

B213-C1-1	&	5.87	$\pm$	0.01	&	0.68	$\pm$	0.02	&	0.49	$\pm$	0.10	&	5.99	$\pm$	0.02	&	0.35	$\pm$	0.04	&	0.11	$\pm$	0.02	&	5.95	$\pm$	0.01	&	0.57	$\pm$	0.04	&	0.18	$\pm$	0.04	&	5.99	$\pm$	0.07	&	0.57	$\pm$	0.14	&	0.03	$\pm$	0.01	\\
B213-C2-1	&	7.08	$\pm$	0.01	&	0.30	$\pm$	0.02	&	0.20	$\pm$	0.04	&	7.06	$\pm$	0.02	&	0.13	$\pm$	0.05	&	0.04	$\pm$	0.01	&	7.03	$\pm$	0.04	&	0.49	$\pm$	0.09	&	0.07	$\pm$	0.01	&				&				&	<0.05			\\
B213-C5-1	&	6.51	$\pm$	0.01	&	0.66	$\pm$	0.03	&	0.35	$\pm$	0.07	&				&				&				&				&				&	<0.05			&				&				&	<0.05			\\
B213-C6-1	&	6.87	$\pm$	0.01	&	0.83	$\pm$	0.02	&	0.30	$\pm$	0.06	&				&				&				&	7.06	$\pm$	0.02	&	0.41	$\pm$	0.05	&	0.09	$\pm$	0.02	&				&				&	<0.03			\\
%B213-C7-1	&	6.99	$\pm$	0.03	&	0.55	$\pm$	0.05	&	0.20	$\pm$	0.04	&	6.82	$\pm$	0.03	&	0.17	$\pm$	0.06	&	0.07	$\pm$	0.01	&				&				&				&				&				&				\\
B213-C10-1	&	6.85	$\pm$	0.02	&	0.29	$\pm$	0.05	&	0.22	$\pm$	0.04	&				&				&				&				&				&	<0.11			&				&				&	<0.07			\\
B213-C12-1	&	6.81	$\pm$	0.02	&	0.31	$\pm$	0.03	&	0.24	$\pm$	0.05	&				&				&				&				&				&	<0.08			&				&				&	<0.08			\\
B213-C16-1	&	6.75	$\pm$	0.05	&	0.76	$\pm$	0.09	&	0.08	$\pm$	0.02	&				&				&				&				&				&	<0.04			&				&				&	<0.04			\\
IC348-1	&	8.96	$\pm$	0.01	&	0.81	$\pm$	0.02	&	0.98	$\pm$	0.20	&				&				&				&				&				&	<0.06			&				&				&	<0.04			\\
%IC348-10	&	8.51	$\pm$	0.01	&	0.70	$\pm$	0.02	&	0.47	$\pm$	0.09	&				&				&				&				&				&				&				&				&				\\
L1448-1	&	4.64	$\pm$	0.01	&	0.73	$\pm$	0.03	&	0.65	$\pm$	0.13	&	4.43	$\pm$	0.03	&	0.32	$\pm$	0.06	&	0.15	$\pm$	0.03	&	4.59	$\pm$	0.01	&	0.50	$\pm$	0.02	&	0.32	$\pm$	0.06	&	4.74	$\pm$	0.07	&	0.51	$\pm$	0.17	&	0.03	$\pm$	0.01	\\
1333-C3-1	&	8.55	$\pm$	0.01	&	1.22	$\pm$	0.02	&	1.28	$\pm$	0.26	&				&				&				&	8.78	$\pm$	0.04	&	0.69	$\pm$	0.11	&	0.07	$\pm$	0.01	&				&				&	<0.03			\\
1333-C4-1	&	7.65	$\pm$	0.01	&	1.45	$\pm$	0.01	&	2.27	$\pm$	0.45	&	7.63	$\pm$	0.02	&	1.03	$\pm$	0.04	&	0.40	$\pm$	0.08	&	7.72	$\pm$	0.01	&	1.12	$\pm$	0.01	&	0.80	$\pm$	0.16	&	7.92	$\pm$	0.03	&	1.14	$\pm$	0.06	&	0.15	$\pm$	0.03	\\
1333-C5-1	&	7.64	$\pm$	0.01	&	1.01	$\pm$	0.02	&	1.12	$\pm$	0.22	&				&				&				&				&				&	<0.06			&				&				&	<0.06			\\
1333-C6-1	&	7.50	$\pm$	0.01	&	0.45	$\pm$	0.01	&	1.49	$\pm$	0.30	&				&				&				&				&				&	<0.09			&				&				&	<0.08			\\
1333-C7-1	&	7.64	$\pm$	0.01	&	0.64	$\pm$	0.02	&	0.80	$\pm$	0.16	&				&				&				&	7.66	$\pm$	0.03	&	0.62	$\pm$	0.06	&	0.12	$\pm$	0.02	&				&				&	<0.05			\\
1333-C3-14	&	7.91	$\pm$	0.01	&	0.64	$\pm$	0.01	&	1.32	$\pm$	0.26	&	7.96	$\pm$	0.04	&	0.48	$\pm$	0.08	&	0.15	$\pm$	0.03	&	8.11	$\pm$	0.01	&	0.74	$\pm$	0.03	&	0.24	$\pm$	0.05	&				&				&	<0.03			\\
79-C1-1	&	10.44	$\pm$	0.01	&	0.67	$\pm$	0.03	&	0.45	$\pm$	0.09	&	10.55	$\pm$	0.03	&	0.32	$\pm$	0.08	&	0.06	$\pm$	0.01	&	10.39	$\pm$	0.02	&	0.63	$\pm$	0.03	&	0.22	$\pm$	0.04	&	10.49	$\pm$	0.04	&	0.58	$\pm$	0.08	&	0.09	$\pm$	0.02	\\
%Ori-C3-1	&	10.73	$\pm$	0.01	&	1.20	$\pm$	0.02	&	1.47	$\pm$	0.29	&				&				&				&				&				&				&				&				&				\\
Ori-C2-3	&	8.05	$\pm$	0.02	&	1.32	$\pm$	0.04	&	0.73	$\pm$	0.15	&	8.11	$\pm$	0.05	&	0.33	$\pm$	0.08	&	0.09	$\pm$	0.02	&				&				&	<0.04			&				&				&	<0.04			\\
Ori-C1-2	&	10.96	$\pm$	0.01	&	1.59	$\pm$	0.01	&	2.10	$\pm$	0.42	&	11.00	$\pm$	0.03	&	1.00	$\pm$	0.07	&	0.19	$\pm$	0.04	&				&				&	<0.04			&				&				&	<0.03			\\
B1b-Cal	&	6.49	$\pm$	0.01	&	1.11	$\pm$	0.02	&	1.35	$\pm$	0.27	&	6.63	$\pm$	0.01	&	0.73	$\pm$	0.03	&	0.36	$\pm$	0.07	&	6.55	$\pm$	0.01	&	0.88	$\pm$	0.02	&	0.57	$\pm$	0.11	&	6.68	$\pm$	0.03	&	1.01	$\pm$	0.10	&	0.11	$\pm$	0.02	\\
\noalign{\smallskip}

\noalign{\smallskip}                                      
\hline 
\end{tabular}}
\end{table*}

\section{Physical parameters of the starless cores}
\label{Sec: Physical parameters of the starless cores}

The star forming regions Taurus, Perseus and Orion were observed with Herschel and SCUBA as part of the Gould Belt Survey \citep{Andre2010}, and accurate visual extinction (A$_V$) and dust temperature (T$_d$) maps are available \citep{Malinen2012, Hatchell2005, Lombardi2014, Zari2016}. The angular resolution of the A$_V$-T$_d$ maps ($\sim$36$''$) is  similar to that provided by the 30m telescope at 3mm, allowing a direct comparison of continuum and spectroscopic data. Throughout this paper we adopt the relation between visual extinction and molecular hydrogen column density A$_V\approx$N(H$_2$)$\times$10$^{-21}$ mag \citep{Bohlin1978}.

We adopted the molecular hydrogen densities, n(H$_2$), obtained by \citet{RodriguezBaras2021} towards the selected positions. In that paper we used the line intensities of the observed CS, C$^{34}$S, and $^{13}$CS lines to estimate n(H$_2$) at each of the GEMS positions. CS has been largely used as density and column density tracer in the interstellar medium \citep[e.g.][]{Linke1980, Anglada1996, Shirley2003, Wu2010, Zhang2014, Scourfield2020} %\citep{Linke1980, Zhou1989, Tatematsu1993, Zinchenko1995, Anglada1996, Bronfman1996, Laundhardt1998, Shirley2003, Bayet2009, Wu2010, Zhang2014, Scourfield2020}. 
The lines were fitted using the non-LTE molecular excitation and radiative transfer code RADEX \citep{Tak2007} and the collisional coefficients calculated by \citet{Denis2018}. In addition, they assumed that gas and dust are thermalised, i.e. the  kinetic temperature T$_k$ is equal to the dust temperature derived from far infrared and millimeter observations (TMC~1: \citealp{Fuente2019}; B~213: \citealp{Palmeirim2013}; Perseus: \citealp{Zari2016}; Orion: \citealp{Lombardi2014}). Furthermore, the beam filling factor was assumed to be $\sim$1 towards all positions, i.e., we considered that the emission was extended compared to the IRAM 30m beam size at the observed frequencies, which is a reasonable assumption according to our CS data (see \citealp{Fuente2019, Navarro2020, RodriguezBaras2021}). In the calculations we let n(H$_2$) and  N(CS) vary as free parameters and explore their parameter space following the Monte Carlo Markov Chain (MCMC) methodology with a Bayesian inference approach. In particular, we used the \textit{emcee} \citep{Foreman2012} implementation of the Invariant MCMC Ensemble sampler methods by \citet{Goodman2010}. This method was already used in \citet{Riviere2019} and \citet{Navarro2020}, and allowed us to estimate the molecular hydrogen density as long as the two transitions of CS, J=2$\rightarrow$1 and 3$\rightarrow$2, are detected, which was the case of all the starless cores considered in this work sample. The complete procedure, and the caveats that had to be taken into account, are described in detail in \citet{RodriguezBaras2021}. 

\section{The H$_2$S abundance}
\label{Sec: The H2S abundance}

Sulphur atoms in interstellar ice mantles are expected to preferentially form H$_2$S because of the high hydrogen abundances and the mobility of hydrogen in the ice matrix. Studying H$_2$S abundance in the gas phase is essential to our understanding of the chemical processes that lead to sulphur depletion in these environments. In order to derive accurate H$_2$S column densities, avoiding possible opacity effects, we have observed the  o-H$_2$S 1$_{10}\rightarrow1_{01}$ and o-H$_2^{34}$S 1$_{10}\rightarrow1_{01}$ lines.
The o-H$_2$S line emission has been detected in the whole sample, while its isotopologue, o-H$_2^{34}$S, only in ten out of the 24 sources (three in Taurus, five in Perseus and two in Orion). The line parameters from the line Gaussian fits are shown in Table \ref{Table: Line parameters}. We have determined the molecular column densities of o-H$_2$S and o-H$_2^{34}$S using the RADEX code, the volume densities obtained by \citet{RodriguezBaras2021} and the collisional coefficients calculated by \citet{Dagdigian2020}. The obtained column densities are included in Table \ref{Table: Physical properties and column densities}. 
Uncertainties are considered to be of 20\%, consistently with errors associated to the line measurements, and estimated by propagation in quadrature in any calculation provided.

\begin{table*}
\caption {Starless cores physical properties, measured column densities of the studied compounds and estimated H$_2$S, HDS and D$_2$S abundances. Uncertainties are provided within parentheses. NOTES: $^{(1)}$ Determination with direct line measures. $^{(2)}$ The abundance is estimated by scaling the o-H$_2^{34}$S column density using the 22.5 standard ratio \citep{Gratier2016} in those cases where the isotopologue is available, or by direct determination of the o-H$_2$S column density when the isotopologue is not available. An ortho-to-para ratio of 3 is adopted for H$_2$S, and 2 for D$_2$S.}
\label{Table: Physical properties and column densities}
\centering     
\resizebox{\textwidth}{!}{%    
\begin{tabular}{lllllllllll}\\%l}\\
%\multicolumn{8}{l}{Table 1. GEMS sample} \\ 
\hline\hline
\noalign{\smallskip}
Starless core	&	T$_{kin}$	&	A$_V$	&	n(H$_{2}$)	&	o-H$_{2}$S$^{(1)}$	&	o-H$_{2}^{34}$S	&	HDS	&	o-D$_{2}$S	&	X(H$_{2}$S)$^{(2)}$ & 	X(HDS) &	X(D$_2$S)	\\
	&	(K)	&	(mag)	&	(cm$^{-3}$)	&	(cm$^{-2}$)	&	(cm$^{-2}$)	&	(cm$^{-2}$)	&	(cm$^{-2}$)	&	& 	&	\\
\noalign{\smallskip}
\hline
\noalign{\smallskip}
\multicolumn{11}{c}{{\bf Taurus}} \\
B213-C1-1	&	10.9	&	26.9	&	1.23$\cdot10^{4}$	&	1.14$\cdot10^{14}$	&	1.55$\cdot10^{13}$	&	3.25$\cdot10^{13}$	&	1.60$\cdot10^{12}$	&	1.73$\cdot10^{-08}$	& 1.21$\cdot10^{-09}$	& 8.94$\cdot10^{-11}$	\\
& & & (8.83$\cdot10^{3}$) & (2.58$\cdot10^{12}$) & (3.10$\cdot10^{12}$) & (6.49$\cdot10^{12}$) & (3.68$\cdot10^{11}$) & (3.45$\cdot10^{-09}$) & (2.41$\cdot10^{-10}$)	&	(2.05$\cdot10^{-11}$)\\
B213-C2-1	&	11	&	20.9	&	2.79$\cdot10^{4}$	&	6.65$\cdot10^{12}$	&	8.17$\cdot10^{11}$	&	3.90$\cdot10^{12}$	&	<1.33$\cdot10^{12}$	&	1.17$\cdot10^{-09}$	& 1.87$\cdot10^{-10}$	& $<$9.57$\cdot10^{-11}$	\\
& & & (2.64$\cdot10^{4}$) & (6.14$\cdot10^{11}$) & (1.63$\cdot10^{11}$) & (7.80$\cdot10^{11}$) & & (2.34$\cdot10^{-10}$) & (3.73$\cdot10^{-11}$)	& 	\\
B213-C5-1	&	11.9	&	23.6	&	4.68$\cdot10^{4}$	&	1.49$\cdot10^{13}$	&		&	<2.10$\cdot10^{12}$	&	<1.63$\cdot10^{12}$	&	8.43$\cdot10^{-10}$ & <8.88$\cdot10^{-11}$	&	<1.04$\cdot10^{-10}$	\\
& & & (3.29$\cdot10^{4}$) & (6.70$\cdot10^{11}$) &  & & & (3.78$\cdot10^{-11}$) & &	\\
B213-C6-1	&	10.9	&	22.2	&	8.28$\cdot10^{4}$	&	1.02$\cdot10^{13}$	&		&	1.98$\cdot10^{12}$	&	<7.81$\cdot10^{11}$	&	6.10$\cdot10^{-10}$	 & 8.90$\cdot10^{-11}$	&	<5.28$\cdot10^{-11}$\\
& & & (8.05$\cdot10^{4}$) & (3.07$\cdot10^{11}$) &  & (3.95$\cdot10^{11}$) &  & (1.84$\cdot10^{-11}$) & (1.78$\cdot10^{-11}$)	&	\\
%B213-c7-1	&	11	&	20.2	&	8.04$\cdot10^{3}$	&	4.93$\cdot10^{13}$	&	7.71$\cdot10^{12}$	&		&		&	1.15$\cdot10^{-08}$ & 	&		\\
%& & & (3.98$\cdot10^{3}$) & (4.56$\cdot10^{12}$) & (1.54$\cdot10^{12}$) &  &  & (2.29$\cdot10^{-09}$) & 	&	\\
B213-C10-1	&	11.2	&	20.7	&	4.32$\cdot10^{4}$	&	4.37$\cdot10^{12}$	&		&	<2.73$\cdot10^{12}$	&	<1.19$\cdot10^{12}$	&	2.82$\cdot10^{-10}$ & <1.32$\cdot10^{-10}$	&	<8.59$\cdot10^{-11}$	\\
& & & (5.09$\cdot10^{4}$) & (5.87$\cdot10^{11}$) &  &  &  & (3.78$\cdot10^{-11}$) & 	&	 \\
B213-C12-1	&	10.6	&	22.1	&	1.43$\cdot10^{5}$	&	2.00$\cdot10^{12}$	&		&	<9.32$\cdot10^{11}$	&	<5.23$\cdot10^{11}$	&	1.21$\cdot10^{-10}$ & <4.22$\cdot10^{-11}$	&	<3.55$\cdot10^{-11}$	\\
& & & (1.40$\cdot10^{5}$) & (2.25$\cdot10^{11}$) &  &  &  & (1.36$\cdot10^{-11}$) & 	&	\\
B213-C16-1	&	10.3	&	24.8	&	2.07$\cdot10^{4}$	&	8.57$\cdot10^{12}$	&		&	<4.46$\cdot10^{12}$	&	<3.51$\cdot10^{12}$	&	4.61$\cdot10^{-10}$ & <1.80$\cdot10^{-10}$	&	<2.13$\cdot10^{-10}$	\\
& & & (1.83$\cdot10^{4}$) & (1.07$\cdot10^{12}$) &  &  &  & (5.76$\cdot10^{-11}$) & 	&	\\
\multicolumn{11}{c}{{\bf Perseus}} \\
IC348-C1-1	&	21.7	&	21.8	&	4.49$\cdot10^{4}$	&	4.29$\cdot10^{13}$	&		&	<2.24$\cdot10^{12}$	&	<8.49$\cdot10^{11}$	&	2.62$\cdot10^{-09}$ & <1.03$\cdot10^{-10}$	&	<5.84$\cdot10^{-11}$	\\
& & & (2.54$\cdot10^{4}$) & (8.19$\cdot10^{11}$) &  &  &  & (5.01$\cdot10^{-11}$) & 	&	\\
%IC348-10	&	17.5	&	20.1	&	8.72$\cdot10^{4}$	&	9.26$\cdot10^{12}$	&		&		&		&	6.14$\cdot10^{-10}$	&  & \\
%& & & (5.74$\cdot10^{4}$) & (2.39$\cdot10^{11}$) &  &  &  & (1.58$\cdot10^{-11}$) & 	&	\\
L1448-C1	&	15.1	&	28.4	&	1.01$\cdot10^{4}$	&	1.51$\cdot10^{14}$	&	1.98$\cdot10^{13}$	&	4.88$\cdot10^{13}$	&	2.29$\cdot10^{12}$	&	2.09$\cdot10^{-08}$ & 1.72$\cdot10^{-09}$	&	1.21$\cdot10^{-10}$	\\
& & & (7.08$\cdot10^{3}$) & (4.54$\cdot10^{12}$) & (3.96$\cdot10^{12}$) & (9.76$\cdot10^{12}$) & (5.55$\cdot10^{11}$) & (4.18$\cdot10^{-09}$) & (3.44$\cdot10^{-10}$)	&	(2.93$\cdot10^{-11}$)\\
1333-C3-1	&	17.7	&	30.3	&	9.34$\cdot10^{4}$	&	5.18$\cdot10^{13}$	&		&	1.71$\cdot10^{12}$	&	<5.41$\cdot10^{11}$	&	2.28$\cdot10^{-09}$ & 5.65$\cdot10^{-11}$	&	<2.68$\cdot10^{-11}$	\\
& & & (4.35$\cdot10^{4}$) & (6.28$\cdot10^{11}$) &  & (3.42$\cdot10^{11}$) &  & (2.76$\cdot10^{-11}$) & (1.13$\cdot10^{-11}$)	& 	\\
1333-C3-14	&	15.9	&	26.8	&	1.04$\cdot10^{4}$	&	3.18$\cdot10^{14}$	&	2.82$\cdot10^{13}$	&	4.47$\cdot10^{13}$	&	<2.62$\cdot10^{12}$	&	3.15$\cdot10^{-08}$ & 1.68$\cdot10^{-09}$	&	<1.46$\cdot10^{-10}$ \\
& & & (6.44$\cdot10^{3}$) & (4.63$\cdot10^{12}$) & (5.63$\cdot10^{12}$) & (8.94$\cdot10^{12}$) &  & (6.31$\cdot10^{-09}$) & (3.33$\cdot10^{-10}$)	&	\\
1333-C4-1	&	16.8	&	39.9	&	5.76$\cdot10^{4}$	&	2.72$\cdot10^{14}$	&	2.90$\cdot10^{13}$	&	5.49$\cdot10^{13}$	&	4.49$\cdot10^{12}$	&	2.18$\cdot10^{-08}$ & 1.38$\cdot10^{-09}$	&	1.69$\cdot10^{-10}$	\\
& & & (3.70$\cdot10^{4}$) & (1.87$\cdot10^{12}$) & (5.79$\cdot10^{12}$) & (1.10$\cdot10^{13}$) & (8.97$\cdot10^{11}$) & (4.36$\cdot10^{-09}$) & (2.75$\cdot10^{-10}$)	& (3.37$\cdot10^{-11}$)	\\
1333-C5-1	&	19.2	&	27.5	&	8.65$\cdot10^{4}$	&	3.63$\cdot10^{13}$	&		&	<2.02$\cdot10^{12}$	&	<8.97$\cdot10^{11}$	&	1.76$\cdot10^{-09}$ & <7.34$\cdot10^{-11}$	&	<4.89$\cdot10^{-11}$	\\
& & & (4.90$\cdot10^{4}$) & (6.08$\cdot10^{11}$) &  &  &  & (2.95$\cdot10^{-11}$) & 	&	\\
1333-C6-1	&	19.9	&	16.1	&	4.74$\cdot10^{4}$	&	4.53$\cdot10^{13}$	&		&	<1.92$\cdot10^{12}$	&	<8.94$\cdot10^{11}$	&	3.75$\cdot10^{-09}$ & <1.19$\cdot10^{-10}$	&	<8.33$\cdot10^{-11}$	\\
& & & (3.18$\cdot10^{4}$) & (1.89$\cdot10^{12}$) &  &  &  & (1.56$\cdot10^{-10}$) & 	&	\\
1333-C7-1	&	17.3	&	17.6	&	1.35$\cdot10^{5}$	&	1.13$\cdot10^{13}$	&		&	2.20$\cdot10^{12}$	&	<4.11$\cdot10^{11}$	&	8.56$\cdot10^{-10}$ & 1.25$\cdot10^{-10}$	&	<3.50$\cdot10^{-11}$	\\
& & & (8.90$\cdot10^{4}$) & (3.09$\cdot10^{11}$) &  & (4.39$\cdot10^{11}$) &  & (2.34$\cdot10^{-11}$) & (2.50$\cdot10^{-11}$)	& \\
79-C1-1	&	15.4	&	19.9	&	1.93$\cdot10^{4}$	&	4.09$\cdot10^{13}$	&	3.71$\cdot10^{12}$	&	1.89$\cdot10^{13}$	&	4.19$\cdot10^{12}$	&	5.59$\cdot10^{-09}$ & 9.50$\cdot10^{-10}$	&	3.16$\cdot10^{-10}$	\\
& & & (1.86$\cdot10^{4}$) & (4.63$\cdot10^{12}$) & (5.63$\cdot10^{12}$) & (8.94$\cdot10^{12}$) & (2.62$\cdot10^{12}$) & (1.13$\cdot10^{-10}$) & (1.90$\cdot10^{-10}$)	&	(6.32$\cdot10^{-11}$) \\
B1b	&	11.9	&	76.0	&	3.15$\cdot10^{5}$	&	2.69$\cdot10^{13}$	&	5.20$\cdot10^{12}$	&	1.40$\cdot10^{13}$	&	1.20$\cdot10^{12}$	&	2.05$\cdot10^{-09}$ & 1.84$\cdot10^{-10}$	& 2.37$\cdot10^{-11}$	 \\
& & & (3.00$\cdot10^{5}$) & (3.75$\cdot10^{11}$) & (1.04$\cdot10^{12}$) & (2.80$\cdot10^{12}$) & (2.41$\cdot10^{11}$) & (4.11$\cdot10^{-10}$) & 	(3.68$\cdot10^{-11}$) &	(4.75$\cdot10^{-12}$)\\
\multicolumn{11}{c}{{\bf Orion}} \\
%Ori-c3-1	&	21.4	&	129.7	&	9.67$\cdot10^{4}$	&	5.33$\cdot10^{13}$	&		&		&		&	5.47$\cdot10^{-10}$	&  & \\
%& & & (4.73$\cdot10^{4}$) & (7.08$\cdot10^{11}$) &  &  &  & (7.28$\cdot10^{-12}$) & 	&	\\
ORI-C1-2	&	18.1	&	173	&	3.43$\cdot10^{4}$	&	4.28$\cdot10^{14}$	&	2.75$\cdot10^{13}$	&	<3.74$\cdot10^{12}$	&	<1.49$\cdot10^{12}$	&	4.77$\cdot10^{-09}$	 & <2.16$\cdot10^{-11}$	&	<1.30$\cdot10^{-11}$\\
& & & (1.95$\cdot10^{4}$) & (2.18$\cdot10^{12}$) & (5.50$\cdot10^{12}$) &  &  & (9.54$\cdot10^{-10}$) & 	&	\\
ORI-C2-3	&	22.8	&	56.5	&	2.61$\cdot10^{4}$	&	8.09$\cdot10^{13}$	&	3.38$\cdot10^{12}$	&	<3.64$\cdot10^{12}$	&	<1.80$\cdot10^{12}$	&	1.80$\cdot10^{-09}$	 & <6.44$\cdot10^{-11}$	& <4.77$\cdot10^{-11}$	\\
& & & (1.72$\cdot10^{4}$) & (3.20$\cdot10^{12}$) & (6.77$\cdot10^{11}$) &  &  & (3.59$\cdot10^{-10}$) & &	\\
\noalign{\smallskip}                                      
\hline 
\end{tabular}}
\end{table*}

\begin{figure*}
\includegraphics[scale=.43]{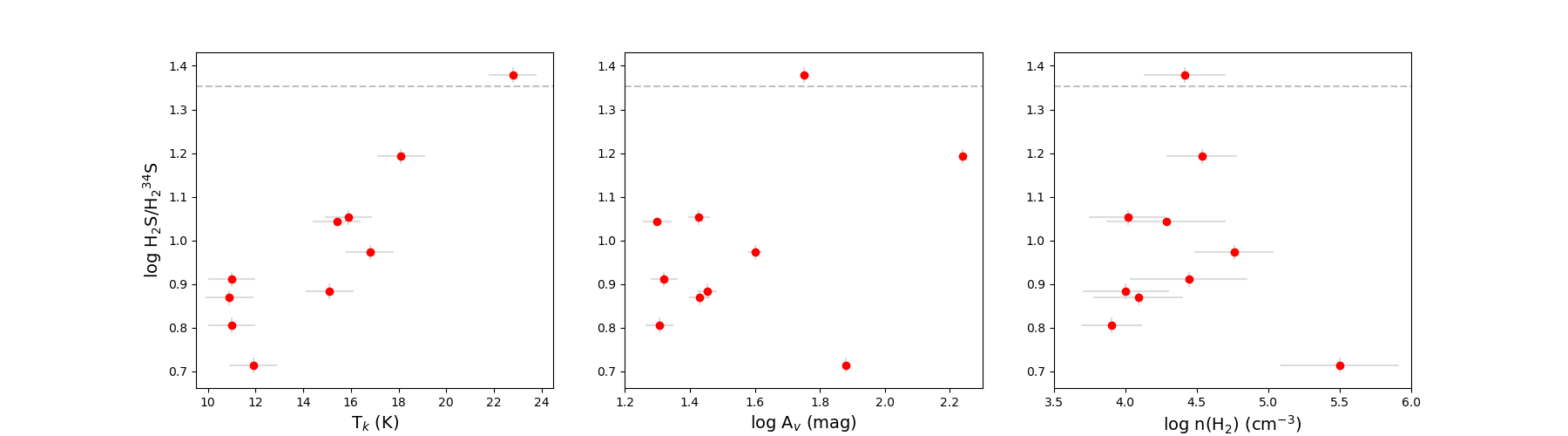}
\caption{Relation of the H$_2$S/H$_2^{34}$S column density ratio to cloud physical parameters: kinetic temperature (left), extinction (middle), and molecular hydrogen density (right). The grey dashed line indicates the value of the standard isotopic value expected for the H$_2$S/H$_2^{34}$S ratio \citep[22.5,][]{Gratier2016}.}
\label{Fig: H2SH234S ratio vs phys prop}
%\vspace{-0.1cm}
\end{figure*}

Fig. \ref{Fig: H2SH234S ratio vs phys prop} presents the relationship between the o-H$_2$S and o-H$_2^{34}$S column density ratio and the main physical parameters of the starless cores (kinetic temperature, extinction and molecular hydrogen density). A clear direct relation with temperature can be observed (with a Pearson correlation coefficient of 0.91), where the standard isotopic value expected for the H$_2$S/H$_2^{34}$S ratio \citep[22.5,][]{Gratier2016} is only reached at the highest temperatures observed in our starless cores. This apparent correlation is the result of the fact that the H$_2$S 1$_{10}\rightarrow1_{01}$ line is optically thick in the dense environments we are considering. In this regime, the fitting to the o-H$_2$S line only provides a lower limit to its column density, and therefore to the H$_2$S/H$_2^{34}$S ratio. To avoid these opacity effects in the analysis of the o-H$_2$S emission and quantities related, in those starless cores where o-H$_2^{34}$S emission is detected we will estimate the o-H$_2$S column density by scaling the o-H$_2^{34}$S column density using the 22.5 standard ratio  \citep{Gratier2016}. Proper distinction will be made between both cases, and it should be taken into account that, when directly measured o-H$_2$S is considered, these column densities and corresponding abundances could be underestimated. 

\begin{figure*}
\includegraphics[scale=.43]{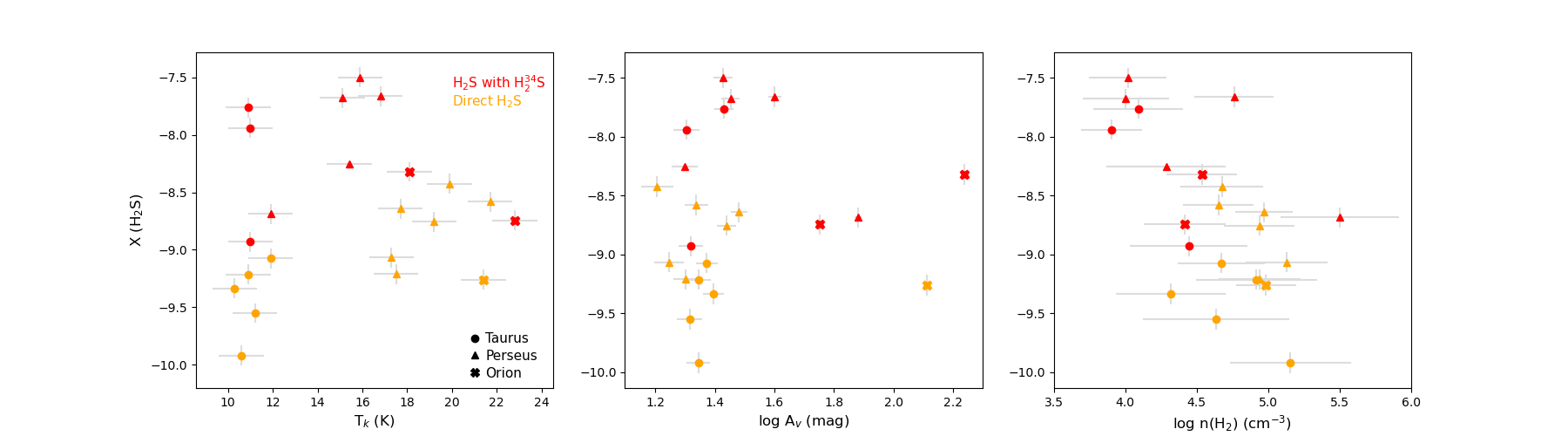}
\caption{Relation of the H$_2$S molecular abundance to cloud physical parameters: kinetic temperature (left), extinction (middle), and molecular hydrogen density (right). The dataset is colour-coded according to the method applied for the estimation of the abundance, and the symbol shape indicate the molecular cloud of the points, as shown in the legends. } %Error bars are always plotted for both axes, although in many cases are smaller than the marker.}
\label{Fig: H2S abundance vs phys prop}
%\vspace{-0.1cm}
\end{figure*}

In the case of molecules formed on the ice surface, the ortho-to-para (OPR) ratio might differ from the thermal equilibrium value. Indeed, laboratory experiments reported by \citet{Hama2018} showed that the OPR of desorbed H$_2$O molecules is close to the statistical value of 3 even at very low temperatures. This  would be the case of H$_2$S molecules that are mainly formed on the grain surfaces as long as subsequent gas-phase chemistry has no time to change it. \citet{Sipila2015} modeled the spin-state chemistry of H$_2$O in starless cores and found that at late times ($\sim$10$^6$ yr), the OPR would be $\sim$1.6. We determined the H$_2$S abundances adopting OPR=3. The thermal equilibrium value of the OPR of H$_2$S is $\sim$1.8 at $\sim$12 K. Therefore, we consider that the uncertainty due to the assumed OPR is less than a factor of 1.7. The relations between the obtained o-H$_2$S abundances and the physical parameters of the starless cores are shown in Fig. \ref{Fig: H2S abundance vs phys prop}. This figure can be directly compared with Fig. 14 by \citet{RodriguezBaras2021}, where analog relations are presented for the complete GEMS observations. The trends and distributions obtained in this case for the starless cores confirm those observed in the mentioned reference. Under the physical conditions prevailing in dark clouds, the main driver of the H$_2$S abundance behaviour seems to be the molecular hydrogen abundance, where a strong anti-correlation is observed. \citet{RodriguezBaras2021} already considered that optical depth effects were not expected to bias the observed anti-correlation with n(H$_2$), and this is confirmed by the same trend shown by starless cores where the H$_2$S abundance is derived from H$_2^{34}$S detections.  Particularly high values of the H$_2$S abundance in a few starless cores could be due to the strong influence of the star formation activity in their surroundings: this is the case, for example, of B213-C1-1, located towards the northern region of Taurus, with a higher stellar formation activity and thus possible influence of outflows, or 1333-C3-14, in Perseus, classified as a protostar. Non-thermal desorption of H$_2$S driven by the enhanced cosmic-ray flux and shocks associated with bipolar outflows are most likely the cause of this behaviour.

\section{Deuterated compounds: HDS and D$_2$S }
\label{Sec: Deuterated compounds: HDS and D2S}

One of the options to address the unsolved questions about the H$_2$S formation and evolution, and the sulphur chemistry in general, is the study of its deuterium fraction through the analysis of the emission of its first deuterated compound, HDS, and its second deuterated compound, D$_2$S. HDS was first observed by \citet{vanDishoeck1995} towards the cold star-forming core IRAS 16293, and after that some upper limits have been measured in several hot cores \citep{Hatchell1999, Crockett2014}. The first detection of D$_2$S was made by \citet{Vastel2003} in NGC~1333~IRAS~4A and B1b, with some upper limits measured in a wider sample of Class 0 sources and dense cores.    

We have detected HDS emission in ten of our starless cores, and D$_2$S, in five. In the case of the Taurus cloud, HDS is detected in three starless cores (B213-C1-1, B213-C2-1, B213-C6-1), while D$_2$S is only located in one (B213-C1-1). These three cores are located in the northern zone of the filament, which is characterised by a higher star formation activity compared to the southern part \citep{Hacar2013}. High deuteration fractions of  H$_2$CS have also been measured in these northern cores by \citet{Esplugues2022}. Regarding Perseus, HDS is detected in seven starless cores (L1448-1, 1333-C3-1, 1333-C4-1, 1333-c7-1, 1333-C3-14, 79-C1-1, B1b) and D$_2$S is detected in four (L1448-1, 1333-C4-1, 79-C1-1, and B1b). IC348 is the only Perseus filament where no deuterated H$_2$S is detected, probably due to the fact that it is located in a warmer environment \citep{Knee2000, Plunkett2013}. There is no detection of HDS or D$_2$S in the Orion starless cores. 

HDS and o-D$_2$S line parameters from Gaussian fits are included in Table \ref{Table: Line parameters}, as well as the intensity upper limits in the case of non-detections. Column densities are derived following the same method explained in Sect. \ref{Sec: The H2S abundance}, using the RADEX code and HDS and o-D$_2$S collisional coefficients by \citet{Dagdigian2022}. Obtained values are included in Table \ref{Table: Physical properties and column densities}. Upper limits are derived for those sources where the corresponding emission is not detected. Abundances are estimated for both HDS and D$_2$S, considering in the case of o-D$_2$S an ortho-to-para (OPR) ratio of 2, and are also included in Table \ref{Table: Physical properties and column densities}.

\subsection{Relation with cloud physical parameters}
\label{Subsec: Deut relation with physical parameters}

\begin{figure*}
\includegraphics[scale=.48]{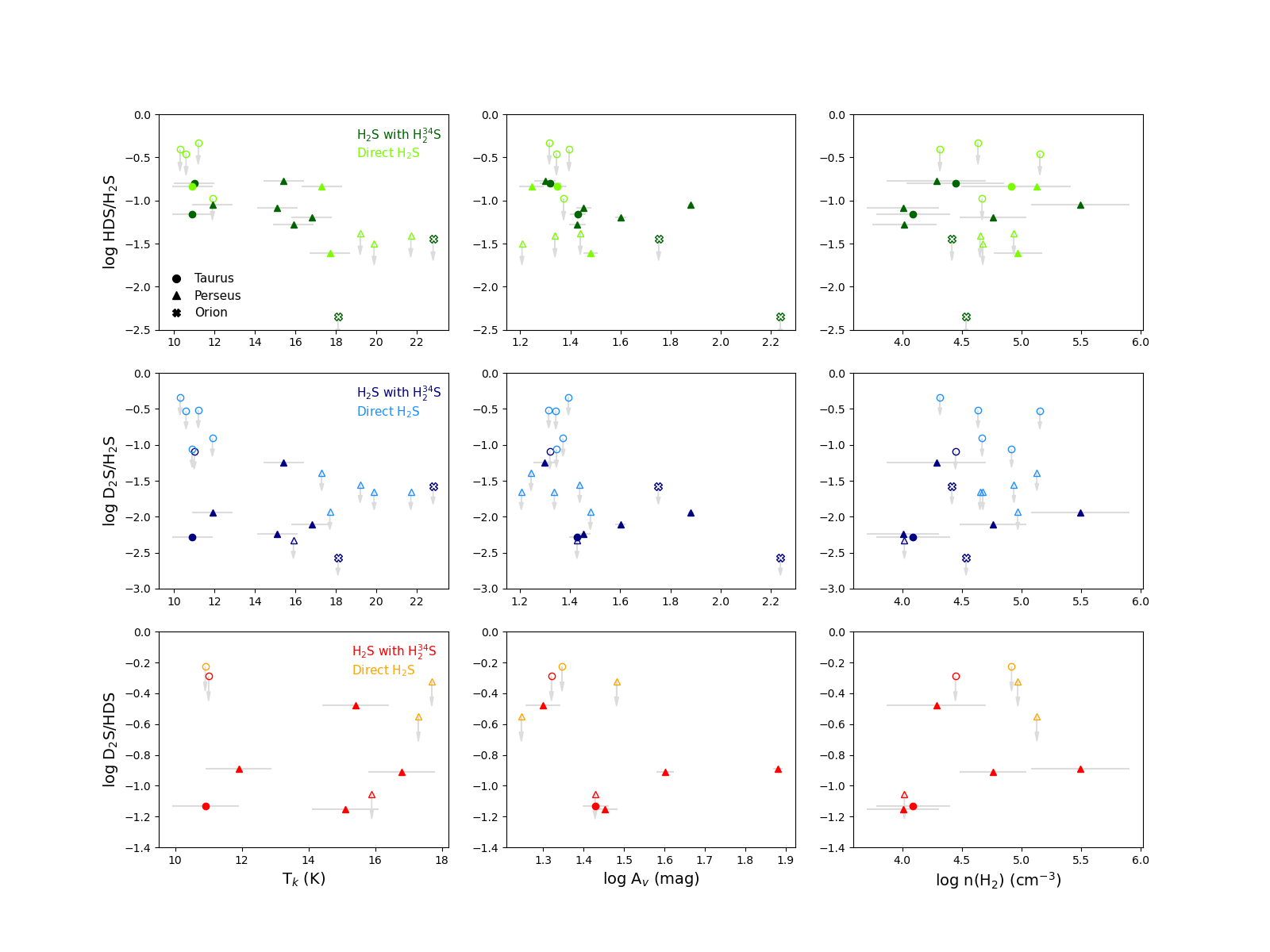}
\caption{Relation between HDS/H$_2$S (top row), D$_2$S/H$_2$S (middle row), and D$_2$S/HDS (bottom row) abundance ratios to starless cores physical parameters: kinetic temperature (left), extinction (middle), and molecular hydrogen density (right). The dataset is colour-coded in each row according to the way of estimation of H$_2$S column density, as indicated in the legends. Different symbols indicate the molecular cloud of the points, as indicated in the legend of the first panel. Filled symbols correspond to direct detections of HDS (top row) or D$_2$S (middle row), open symbols correspond to upper limits, respectively. In the case of D$_2$S/HDS (bottom row), filled symbols indicate direct detection of both HDS and D$_2$S, open symbols indicate direct detection of HDS and upper limits of D$_2$S. }
\label{Fig: Deuterated ratios vs phys prop}
%\vspace{-0.1cm}
\end{figure*}

The relations between HDS/H$_2$S, D$_2$S/H$_2$S, and D$_2$S/HDS abundance ratios to starless cores physical parameters are represented in Fig. \ref{Fig: Deuterated ratios vs phys prop}. Distinction is made between the corresponding molecular clouds, the method applied for the derivation of the H$_2$S abundance and between starless cores with direct measurements or with estimated upper limits of HDS or D$_2$S, as indicated. In the case of the D$_2$S/HDS ratio only sources with direct measurements of both compounds or at least direct measurement of HDS are represented, as the cases with both upper limits introduced an excess of degeneracy. 

Regarding the clouds, single deuteration values (i.e. HDS/H$_2$S abundance ratios) seem to be slightly higher in Taurus than in Perseus, although this should be considered with caution since several of Taurus values are upper limits. One of the Orion starless cores, Ori-C1-2, shows an upper limit to the HDS/H$_2$S ratio lower than the other, with a value more than one order of magnitude lower than those measured in Taurus and Perseus. However, we cannot extrapolate this behaviour to the starless cores located in Orion A. \citet{Esplugues2022} found large differences, about one order of magnitude, in the deuterium fractions of H$_2$CS towards the starless cores located in Orion.

All in all, a decreasing relation between H$_2$S single deuteration and kinetic temperature is observed for the whole sample. This decreasing relation is expected, since low temperatures favor higher values of the D/H ratio. In particular, we have not any detections in positions with T$>$18~K, which confirms that CO depletion is an important parameter for deuteration. The HDS/H$_2$S abundance remains constant with a large scatter for T$<$18~K. Interestingly enough, no trend is observed in the relation between HDS/H$_2$S abundance ratios and molecular hydrogen density, n(H$_2$) within our uncertainties.  The lack of correlations of HDS/H$_2$S with molecular hydrogen density is intriguing. As long as the dust temperature is lower than $\sim$18~K, higher densities induce higher CO depletions, which favor the enhancement of H$_{3}^{+}$ abundances and the abundances of deuterated ions H$_2$D$^+$, D$_2$H$^+$, D$_3^+$ in cold gas. Moreover, the deuteration on grain surfaces also depends on the amount of H$_2$D$^+$, D$_2$H$^+$, D$_3^+$ in gas phase since the dissociative electronic recombination of these atoms are producing the D atoms that will eventually accrete on the grain surfaces. The large scatter in the observed values could hinder any trend with the density given the small number of points in our sample. No clear trend is observed in the case of the relation between the HDS/H$_2$S ratio and the extinction, either.

Differences between clouds are less evident in the case of double deuteration values (D$_2$S/H$_2$S abundance ratios). There is only one detection in Taurus, with no significant upper limits in the case of other Taurus objects, which prevents a good comparison with Perseus, and a similar situation with Orion. Analogously, we cannot establish the possible existence of a trend in the relation of this ratio with kinetic temperature or extinction.

The D$_2$S/HDS abundance ratio provides information about the relation between single and double deuteration. We have only 5 detections for the three clouds, which makes it difficult to find any trend. No clear trend is observed in this case of temperature and visual extinction. A hint of trend is observed in the right panel of Fig.~\ref{Fig: Deuterated ratios vs phys prop}, with the D$_2$S/HDS abundance ratio increasing with density. This is consistent with results obtained by \citet{Chacon-Tanarro2019} analysing the emission of deuterated H$_2$CO towards the pre-stellar core L1544.

\subsection{Comparison with other interstellar sources}
\label{Subsec: Deut relation with interstellar sources}

Deuterium fraction is sensitive to physical conditions of the source, such as temperature, CO depletion, H$_2$ density and OPR. Moreover, deuterium fraction in the gas phase is thought to vary along the dynamical evolutionary stages of the star formation process, increasing towards the onset of star formation and decreasing afterwards, due to the heating produced by the newly formed stellar object. Comparison of deuterated emission between different interstellar sources allows to determine the effect of physical conditions and stellar formation evolution on deuterium fraction. 

\begin{figure*}
\centering
\includegraphics[scale=.38]{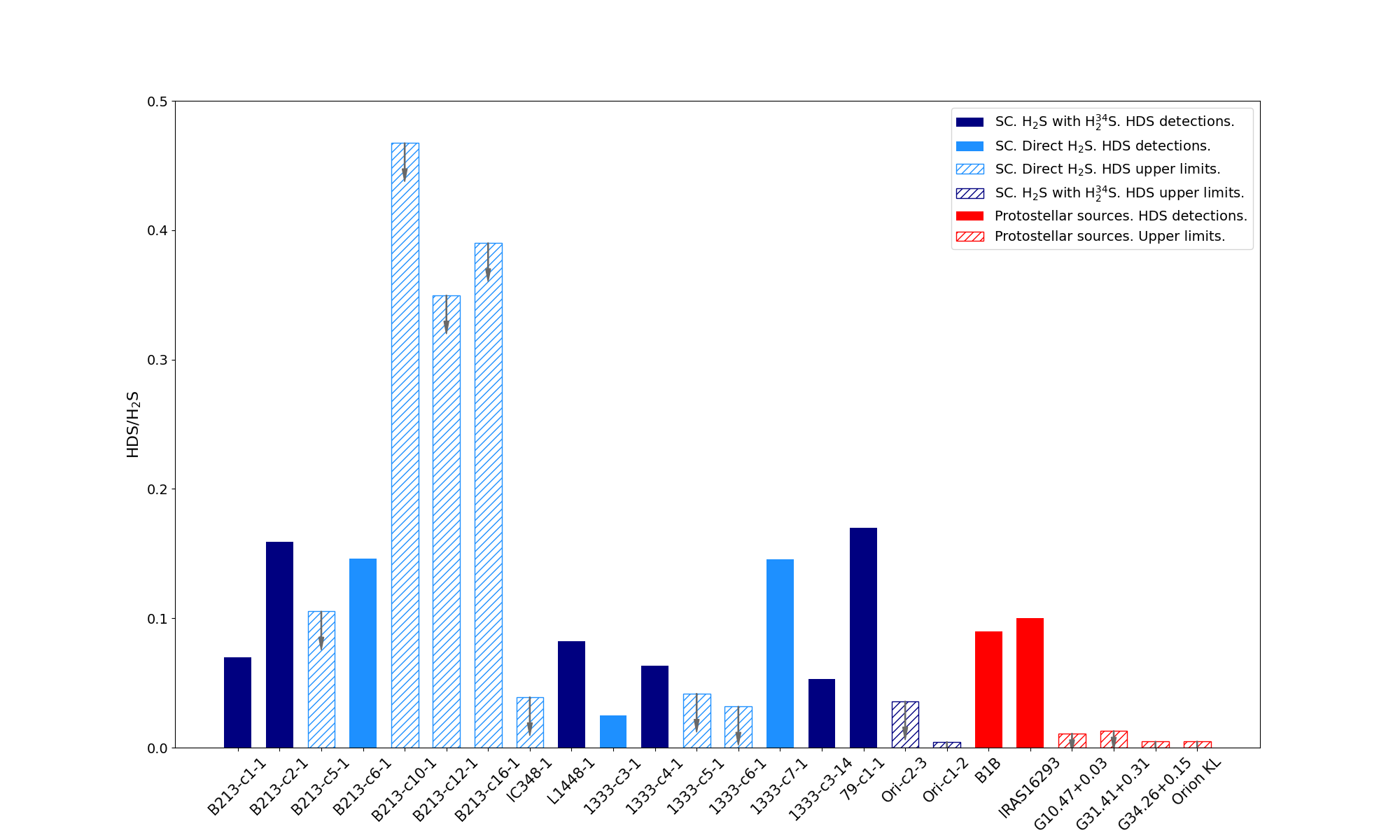}
\caption{Comparison of H$_2$S deuterium fraction, HDS/H$_2$S, between starless cores (blue) and protostellar sources (red). The bar pattern indicates the direct detection (solid) or determination of upper limits (dashed) of the HDS emission. The starless cores dataset is colour-coded according to the method applied for the estimation of the H$_2$S abundance, as shown in the legend. References: Starless cores: This work.B1b: This work. IRAS 16293: \citet{vanDishoeck1995}. G10.47+0.03, G31.41+0.31 and G34.26+0.15: \citet{Hatchell1999}. Orion KL: \citet{Crockett2014}.}
\label{Fig: HDSH2S evolutionary stages}
%\vspace{-0.1cm}
\end{figure*}

Fig. \ref{Fig: HDSH2S evolutionary stages} shows a comparison between the values of the HDS/H$_2$S abundance ratio obtained for our starless cores sample and other sources from the literature. H$_2$S deuterated compounds have been observed in only a few other interstellar sources, and in several cases only upper limits are estimated, but those limits are significant in comparison with values obtained for our sources. Within our sample, it should be noted that no significant difference is found between the HDS/H$_2$S ratio of B1, already classified as a Class 0 source although extremely young \citep{Gerin2015b, Fuente2016, Gerin2017, Fuente2017, Marcelino2018}, and the starless cores of the sample. Starless cores values are also comparable to the one obtained by \citet{vanDishoeck1995} for the hot corino IRAS 16293, where the high temperature of the central region would be thought to produce a decrease of the fraction of deuterated molecules in gas phase. This result, if confirmed, would point again towards the possibility of H$_2$S deuterated compounds mainly formed by surface reactions (see Sec. \ref{Subsec: Deut relation with physical parameters}). If these compounds are subsequently  evaporated due to the increasing temperature produced by the new stellar object, but not utterly processed in the gas phase, deuterium fraction would remain comparable to that of the starless core phase. However, it is also possible that this is the consequence of an observational bias, where observed values of  B1b and IRAS 16293 are still produced by their cold envelopes instead in the compact hot corinos. Interferometric observations with high spatial resolution will be needed to resolve the central region of this kind of sources and confirm or discard this hypothesis. The effect of temperature is indeed observed in the hot massive star-forming clouds G10.47+0.03, G31.41+0.31 and G34.26+0.15, observed by \citet{Hatchell1999}, where only upper limits are measured but are significantly lower than those obtained for the starless cores of our sample. Fig. \ref{Fig: HDSH2S evolutionary stages} also shows that HDS/H$_2$S values of the starless cores sample are clearly higher than the upper limit obtained for Orion KL. This upper limit was derived by \citet{Crockett2014}, who analysed the H$_2$S emission towards the Orion KL hot core, estimating a rotational temperature, T$_{rot}$, of 141 K. They have therefore analysed the H$_2$S emission coming from the inner envelope, where high temperature produces the sublimation of the ice grain mantles built up during the pre-stellar core phase. Traditionally, two explanations have been proposed to explain the differences in the deuteration between hot corinos and hot cores associated with massive star forming regions \citep{Fuente2005, Fuente2014, Ceccarelli2014}. One possibility is that the hot cores associated with massive stars are older and/or significantly denser than those surrounding low-mass objects. In this case, gas phase chemistry could have had the time to reset the deuterium fractions to values close to the cosmic D/H ratio. An alternative explanation is that the temperature of the envelope material accreting onto the high-mass protostars were higher than in the case of low mass stars, already reducing the deuterium fractionation efficiency before the hot core phase. We have observed lower H$_2$S deuterium fractions in the starless cores located in Perseus and Orion than in Taurus. This suggests that the low H$_2$S deuteration in hot cores might indeed be inherited from the pre-stellar phase when the ice mantles are grown. However, our observations towards Orion A are limited to two positions, clearly insufficient to establish firm conclusions. A high sensitive and spatial resolution survey of HDS in  large sample of starless cores in Orion is required to have a deeper insight in the H$_2$ deuteration in Orion.

\begin{figure*}
\includegraphics[scale=.38]{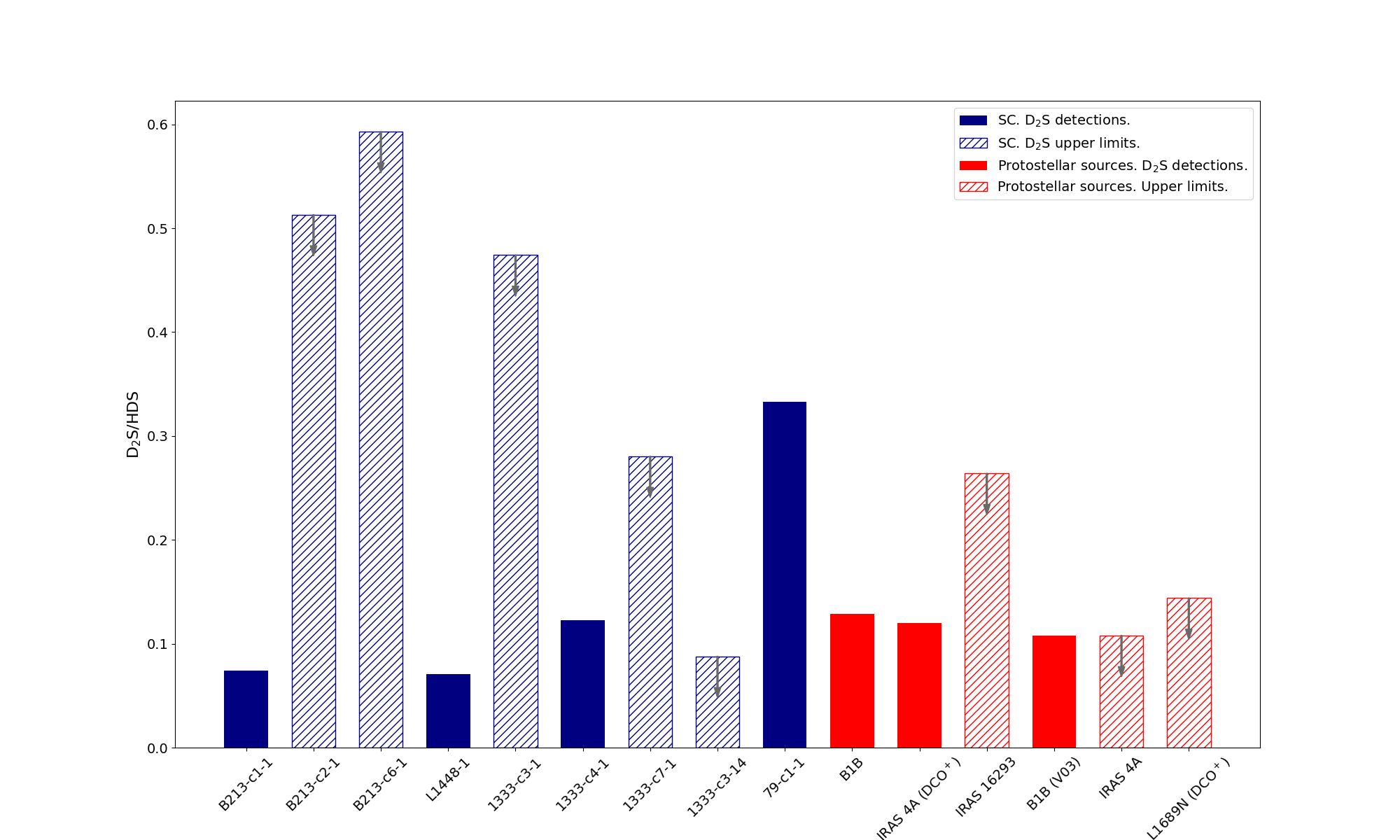}
\caption{Comparison of double deuterium fraction, D$_2$S/HDS, between starless cores (blue) and protostellar sources (red). The bar pattern indicates the direct detection of the ratio lines (solid) or determination of upper limits in the case of D$_2$S emission (dashed), as indicated in the legend. References: Starless cores: This work.B1b: This work and \citet{Vastel2003} (marked in the plot as V03). Rest of protostellar sources: \citet{Vastel2003}.}
\label{Fig: D2SHDS evolutionary stages}
\end{figure*}

Analogously, a comparison between the D$_2$S/HDS abundance ratio values obtained for our sample and other interstellar sources from the literature is shown in 
Fig.\ref{Fig: D2SHDS evolutionary stages}. In this case only \citet{Vastel2003} provide observations of double deuterated H$_2$S for five sources, with two detections and three upper limits. The value obtained in this work for B1b is in good agreement with that obtained by these authors for the same source. Although the number of D$_2$S detections is limited and therefore establishing a firm conclusion is risky, our data may suggest that the D$_2$S/HDS abundance ratio is increasing along the starless core evolution, becoming maximum close to the Class 0 phase.

\subsection{Comparison with other molecules}
\label{Subsec: Comparison with other molecules}

The deuterium fraction is a powerful tool to study the evolution of material during the star and planetary formation process. As an example, the observation and modelling of water and its deuterated isotopologues has established that a substantial fraction of the water in the Solar System has been inherited from the prestellar core, where the Sun was formed \citep{Cleeves2014, vanDishoeck2021}. The conclusive evidence of this was found in the high abundances of the doubly deuterated water, highlighting how powerful it is to use deuteration as a probe of inheritance. Moreover, molecules with multiple deuteration provide crucial constraints not only to the deuteration processes involved, but also to the formation of the main species.

In order to be able to compare between D/H ratios observed for different molecules, it has to be taken into account that the deuterium fraction of a molecule does not directly correspond to the abundance ratio between the deuterated and the main isotopologues, as presented in Appendix B of \citet{Manigand2019} and Appendix C of \citet{Drozdovskaya2022}. When the chemical group containing D isotopes has at least two bonds to H or D atoms, it is not possible to distinguish between the different arrangements of H and D. The probability of a D atom replacing a H atom at a specific location in a certain functional group (-XH$_n$) is a statistically independent event. The number of indistinguishable combinations with \textit{i} deuterium atoms at \textit{n} potential sites in a specific functional group is the statistical correction that has to be applied to the abundance ratio of two isotopologs to obtain the molecular deuteration fraction, i.e. the D/H ratio, of a molecule. The relation between the abundance ratio and the D/H ratio is:

\begin{equation}
	    \frac{XH_{n-i}D_i}{XH_n} = \binom{n}{i}\, \left(\frac{D}{H}\right)^i_{XH_n},
\end{equation}

as presented in \citet{Manigand2019}, where \textit{n} is the number of valence of the X group, \textit{i} is the number of D attached to the X group, and $\binom{n}{i} = \frac{n!}{i!(n-i)!}$ is the number of arrangements of \textit{i} into \textit{n}. When the D/H ratio of an isotopologue is considered relative to another isotopologue that already contains \textit{j} deuterium atoms in that functional group, instead of the isotopologue without deuterium, then this leads to the relation:

\begin{equation}
	    \frac{XH_{n-i}D_i}{XH_{n-j}D_j} =\frac{\binom{n}{i}}{\binom{n}{j}}\,\left(\frac{D}{H}\right)^{i-j}_{XH_{n-j}D_j}, i>j\geq0.
\end{equation}

The different relations between abundance ratio and D/H ratio for the isotopologues considered in this paper are provided in Table \ref{Table: Statistical corrections}.

\begin{table}
\caption{Relations between the abundance ratio and the D/H ratio.}
\label{Table: Statistical corrections}
\centering
\begin{tabular}{rcl}\\%ll}\\%l}\\%l}\\
%\multicolumn{8}{l}{Table 1. GEMS sample} \\ 
\hline\hline
\noalign{\smallskip}
Abundance ratio &  & D/H ratio prescription \\% & N$_{\rm o}$ & N$_{\rm v}$ \\%& n(H$_2$)\\%& CD(CS)\\
\hline
\noalign{\smallskip}
HDS/H$_2$S & = & 2 (D/H)\\
D$_2$S/HDS & = & 1/2 (D/H)$_\mathrm{HDS}$\\
\hline
\noalign{\smallskip}
HDCS/H$_2$CS & = & 2 (D/H)\\
D$_2$CS/HDCS & = & 1/2 (D/H)$_\mathrm{HDCS}$\\
\hline
\noalign{\smallskip}
HDO/H$_2$O & = & 2 (D/H)\\
D$_2$O/HDO & = & 1/2 (D/H)$_\mathrm{HDO}$\\
\hline
\noalign{\smallskip}
HDCO/H$_2$CO & = & 2 (D/H)\\
D$_2$CO/HDCO & = & 1/2 (D/H)$_\mathrm{HDCO}$\\
\hline
\noalign{\smallskip}                          
CH$_2$DOH/CH$_3$OH & = & 3 (D/H) \\
CH$_3$OD/CH$_3$OH & = & (D/H) \\
CHD$_2$OH/CH$_2$DOH & = & (D/H)$_\mathrm{CHD_2OH}$ \\
\hline
\noalign{\smallskip}
cC$_3$HD/cC$_3$H$_2$ & = & 2 (D/H)\\
cC$_3$D$_2$/cC$_3$HD & = & 1/2 (D/H)$_\mathrm{cC_3HD}$\\

\hline
\noalign{\smallskip}
\end{tabular}
\end{table}

\begin{figure*}
\centering
\includegraphics[scale=.34]{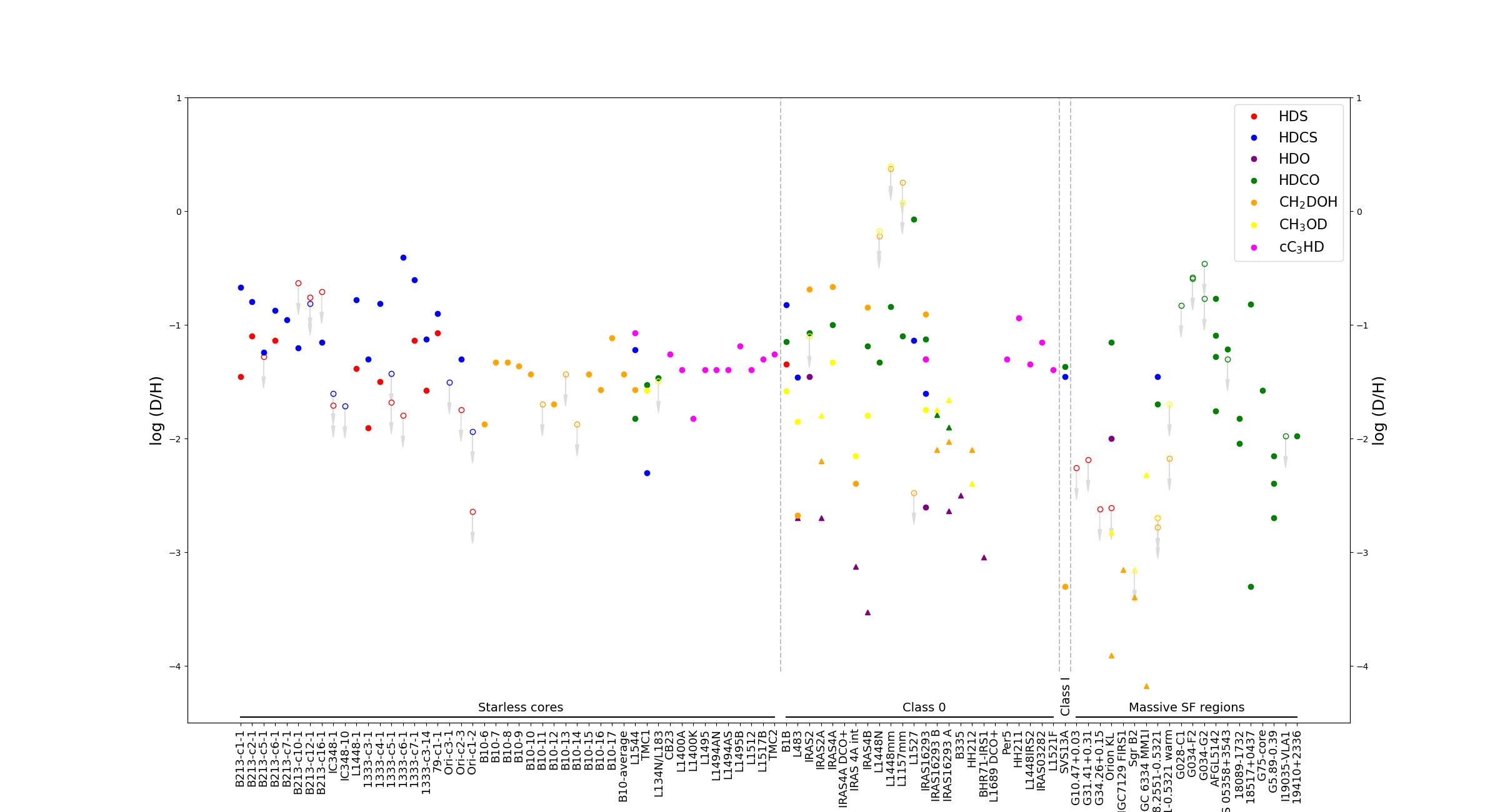}
\caption{Comparison between D/H ratios of simply deuterated molecules, including the respective statistical corrections (see text), for several early stage star-forming sources. Data are colour-coded corresponding to the different molecules, shown in the legend. Filled symbols represent direct detections of the corresponding deuterated molecule, open symbols represent upper limits. Circles represent data obtained with single-dish observations, triangles represent data obtained with interferometric observations. Grey dashed lines are a guide to distinguish different star-forming stages.\\
\textbf{NOTES:} References per source: Starless cores GEMS sample: This work and \citet{Esplugues2022}. B10-6 to B10-17 starless cores in Taurus: \citet{Ambrose2021}. L1544: \citet{Chacon-Tanarro2019, Spezzano2022b, Giers2022}. TMC1: \citet{Minowa1997, Turner2001}. L134N/L183: \citet{Turner2001}. CB23, L1400A, L1400K, L1495, L1495AN, L1495AS, L1495B, L1512, L1517B, TMC2: \citet{Chantzos2018}.B1b: This work and \citet{Marcelino2005, Turner2001}. L483: \citet{Agundez2019, Jensen2021}. IRAS2/IRAS2A: \citet{Coutens2014a, Liu2011, Parise2006, Taquet2013b}. IRAS4A/IRAS4A DCO$^+$/IRAS4A interferometry: \citet{Parise2006, Taquet2013b, Taquet2019, Vastel2003}. IRAS4B: \citet{Parise2006, Persson2014}. L1448N/L1448mm: \citet{Parise2006}. L1157mm: \citet{Parise2006}. L1527: \citet{Parise2006, Sakai2009b, Yoshida2019}. IRAS16293: \citet{Butner2007, Coutens2012, Drozdovskaya2018, Parise2006, vanDishoeck1995, Vastel2003, Chantzos2018}. IRAS16293A: \citet{Drozdovskaya2022, Manigand2020, Persson2013}. IRAS16293 B: \citet{Drozdovskaya2022, Jorgensen2018, Persson2018}. B335: \citet{Jensen2021}. HH212: \citet{Bianchi2017b, Taquet2019}. BHR71-IRS1: \citet{Jensen2019}. L1689 DCO$^+$: \citet{Vastel2003}. Per5, HH211, L1448IRS2, IRAS03282, L1521F: \citet{Chantzos2018}. SVS13A: \citet{Bianchi2017a, Bianchi2019b}. G10.47+0.03, G31.41+0.31, G34.26+0.15: \citet{Hatchell1999}. Orion KL: \citet{Bergin2002, Crockett2014, Lis2002, Taquet2019, Turner1990}. NGC7129 FIRS2: \citet{Fuente2014}. Sgr B2: \citet{Belloche2016}. NGC 6334 MM1I: \citet{Bogelund2018}. G328.2551-0.5321/G328.2551-0.5321 warm: \citet{Bouscasse2022}. AFGL5142, IRAS~05358+3543, G5.89-0.39: \citet{Zahorecz2017, Zahorecz2021}. G028-C1, G034-F2, G034-G2, 18089-1732, 18517+0437, G75-core, I19035-VLA1, 19410+2336: \citet{Zahorecz2021}.  }
\label{Fig: DH molecules and sources}
\end{figure*}

Fig. \ref{Fig: DH molecules and sources} shows the D/H ratios, including the respective statistical corrections, of simply deuterated isotopologues of H$_2$S, H$_2$CS, H$_2$CO, c-C$_3$H$_2$, and CH$_3$OH measured in our starless cores sample, as well as other starless cores and young star-forming sources published in the literature. First of all, this allows a direct comparison between the D/H values measured for H$_2$S in this work  with the D/H ratio derived from the HDCS abundances reported by \citet{Esplugues2022} in the same sample. This comparison reveals that the D/H is systematically lower in H$_2$S than in H$_2$CS in starless cores. One could think that this difference is related to the different formation routes for these two molecules. While thioformaldehyde (H$_2$CS) can be formed through gas-phase reactions and on grain surfaces \citep{Esplugues2022}, the formation of H$_2$S, and probably its deuteration, should occur only on the grain surfaces \citep{Oba2019}. Alternatively, lower deuteration of H$_2$S might also be related to an earlier formation time of this molecule in the collapsing process of the starless core, with higher temperatures and worse conditions for deuteration. It is also interesting to compare the deuterium fractions of H$_2$S with that of methanol (CH$_3$OH). \citet{Ambrose2021} observed single-deuterated methanol compound CH$_2$DOH in a sample of starless cores in the B10 region of Taurus. They observed D/H methanol ratios in the same range of values than those that we have obtained for H$_2$S in our sample. Methanol is formed in the cold interior of molecular clouds by successive hydrogenation of CO on the icy grain mantles \citep{Vasyunin2017, Punanova2022} while H$_2$S is more abundant in the cloud surface \citep{Navarro2020}.  One would expect higher values of deuterium fraction in CH$_2$DOH than in H$_2$S.  On the contrary, the two molecules present similar values. Interestingly, CH$_2$DOH with CH$_3$OD present different D/H values, CH$_3$OD being less abundant than CH$_2$DOH. To complete this comparison, we include the D/H ratio of c-C$_3$H$_2$ obtained by  \citet{Chantzos2018} and \citet{Giers2022} in a sample of starless cores and protostars. This molecule is thought to be formed mainly in gas phase. Its D/H ratio is similar to those  of CH$_3$OH and H$_2$S, and lower than H$_2$CS. 

When considering young protostellar objects, classified as Class 0, it is important to differentiate between single-dish and interferometric measurements. Single dish measurements do not allow to discern the origin of the emission within the protostellar envelope, except in the case of multi-transitional studies that are not the usual case.  In Fig. \ref{Fig: DH molecules and sources} we differentiate between those values obtained with single-dish observations and those related to interferometric observations using different symbols. Restricting to single-dish measurements, we observe a wider dispersion of the data, but the D/H values are in the same range as those observed in the case of starless cores. There is a scarcity of HDS detections in this kind of sources, but as seen in Fig. \ref{Fig: HDSH2S evolutionary stages}, they are in good agreement with values obtained in our work for starless cores. Moreover, in these sources there are a certain number of detections of formaldehyde (H$_2$CO), which share the same range of values with those obtained for HDCS, supporting the idea that they may be formed in similar conditions. In the case of methanol, most of the deuterium fraction values obtained both for CH$_2$DOH and CH$_3$OD are lower than those of HDS, HDCS and HDCO, with the exception of four CH$_2$DOH detections obtained by \citet{Parise2006} for IRAS16293, IRAS2, IRAS4A and IRAS4B. Some of these sources also have detections of singly deuterated water, HDO, whose deuterium fraction is smaller than that of any of the other considered molecules. 
While single-dish detections of Class 0 sources share the same range of deuterium fraction values as those obtained for starless cores, interferometric values show a clear decreasing trend with respect to these earlier stage sources. This difference points again to the idea that single-dish observations may be detecting only the emission of the cold envelope, while interferometric observations manage to detect the effect of the increasing temperature of the protostar in the innermost layers of the gas. 

We also compare our results in Fig. \ref{Fig: DH molecules and sources} with values obtained for one Class I source, SVS13A, with detections of deuterated formaldehyde and methanol \citep{Bianchi2017a}, and thioformaldehyde \citep{Bianchi2019b}. HDCO and HDCS deuterium fractions are in the same range of values than those of Class 0 sources. On the contrary, CH$_2$DOH shows a clear decrease with respect to values of younger sources, an effect which cannot be associated to observational techniques since these detections were made with the IRAM 30m telescope. 

Finally, we  observe a general trend (all molecules) of decreasing deuterium fraction values when considering detections of these deuterated molecules in massive star forming regions, an effect that is also in this case even more pronounced in the case of interferometric observations. Again, we consider that we are observing the effect of higher temperatures, creating less favorable environments for deuteration processes.

\begin{figure}
\centering
\includegraphics[scale=.38]{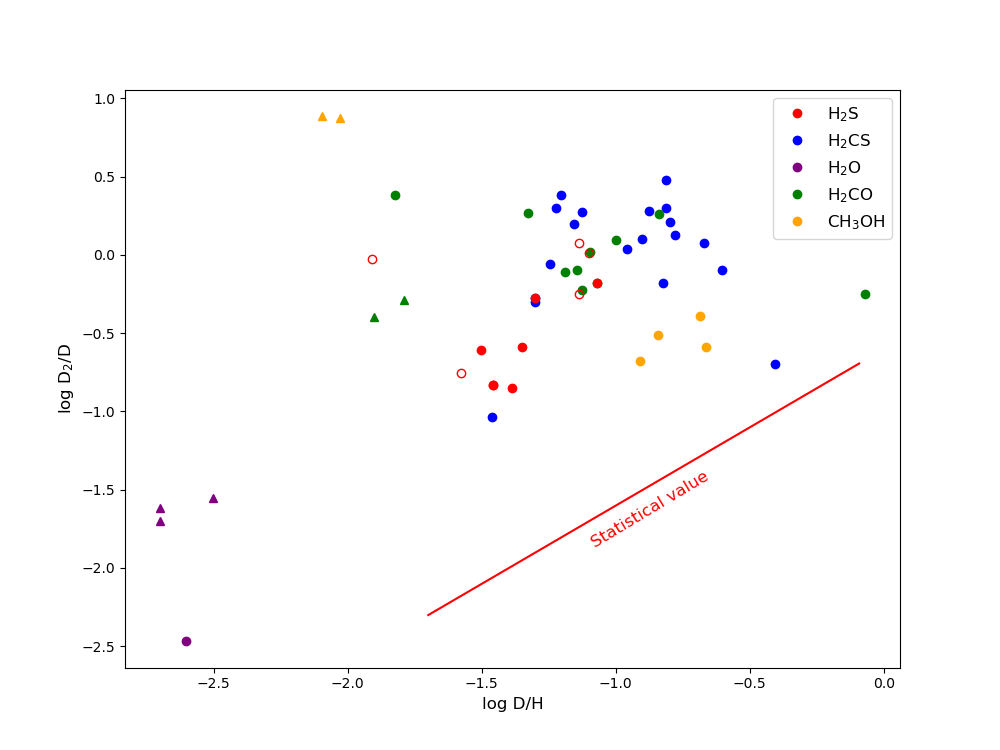}
\caption{D/H ratio of double deuterated molecules as a function of D/H ratio of the corresponding simple deuterated molecules, including statistical corrections in both cases (see text). Only starless cores and Class 0 sources are included, for direct comparison. The dataset is colour-coded according to the corresponding molecule, as indicated in the legend. In the case of methanol, the single deuterated molecule is CH$_2$DOH. Circles represent data obtained with single-dish observations for both ratios, triangles represent data obtained with interferometric observations for both ratios. In the case of H$_2$S, open symbols are upper limits, corresponding to sources of this work where only HDS was directly detected. The red line shows the statistical value if the D atoms were statistically distributed in the molecules formed in the grain surfaces (see text).}
\label{Fig: Double deuteration}
\end{figure}

The study of double deuteration also provides information about the formation and deuteration processes of molecules. In particular, in the case of species forming on the grain surfaces, then the relation D$_2$/D = 0.25 D/H would be hold as long as the distribution of D atoms is purely statistical, meaning just proportional to the D/H ratio, \citep{Ceccarelli2014}. Fig. \ref{Fig: Double deuteration} represents double deuteration fraction, i.e. the abundance ratio of the double deuterated compound to the simple deuterated compound with the corresponding statistical correction, as a function of the simple D/H ratio of the corresponding surface molecule. The statistical relation D$_2$/D = 0.25 D/H (red line in  Fig.\ref{Fig: Double deuteration}) is not fulfilled for any of the considered molecules, a similar result to the one obtained by \citet{Ceccarelli2014}. They provided the plausible explanation that this is the consequence of the D and D$_2$-bearing compounds being formed at different times on the grain surfaces, related to different D/H ratios in the gas. Within this scenario the single-dish detections of methanol, closer to the statistical value, may be related to formation in a short time interval, while the formation of other molecules may occur in larger intervals of times, where the single and double deuterated forms may not have inherited the same D/H atomic ratio. As discussed in next section, new experimental results suggest that these differences could also be related to H-D substitution reactions on the grain surfaces \citep{Oba2019}.

\section{Discussion}
\label{Sec: Discussion}

Single and multiple deuterium fraction provide essential information about the formation and evolution of important molecules of the interstellar medium. This is the case, for example, of molecules formed with an interplay of gas-phase and grain-surface chemistry, such as formaldehyde \citep{Zahorecz2021} and thioformaldehyde. H$_2$CS is formed both in the gas phase \citep[mainly from atomic S,][]{Esplugues2022} and on the surface of dust grains, by addition of H atoms on CS \citep{Spezzano2022b}. Another molecule of interest regarding the grain surface deuteration processes is methanol. In prestellar cores, once CO freezes out onto the surfaces of dust grains, it can either be stored in the icy mantles or react with other elements, particularly atomic hydrogen. In the latter case, CO is first transformed into HCO, then formaldehyde and eventually into methanol. Another example is water, where, as already mentioned in Sec. \ref{Subsec: Comparison with other molecules}, the observation and modelling of the main compound and its deuterated isotopologues established that a substantial fraction of the water in the Solar System has been inherited from the prestellar core, where the Sun was formed \citep{Cleeves2014, vanDishoeck2021}. 

In Fig.~\ref{Fig: DH molecules and sources}, we compare the deuteration (D/H) ratio of H$_2$S with those obtained for H$_2$CS, H$_2$CO, c-C$_3$H$_2$, H$_2$O and CH$_3$OH in different star forming regions. As expected, there is a trend with deuterium fraction being maximum in starless cores and the cold envelope of Class 0 objects, and decreasing in more evolved Class I targets and massive star forming regions. This is a general behaviour shared by all the considered species, nevertheless there are significant differences between species when carrying out a more detailed analysis. In particular, for starless cores, the highest values of D/H are found for HDCS and HDCO while HDS presents lower values similar to those found for deuterated CH$_3$OH  and c-C$_3$H$_2$.

For Class 0 objects, single-dish results (dotted points in Fig. \ref{Fig: DH molecules and sources}) of D/H for H$_2$CO, H$_2$CS, H$_2$S and CH$_3$OH are in the same range as for starless cores. However, if we consider interferometric results (triangles in Fig. \ref{Fig: DH molecules and sources}), we observe a significant decrease in the D/H results with respect to the single-dish observations due to the fact that, in the former case, we are observing only the hot corino. Regardless of the observational technique, single-dish or interferometry, the lowest values of D/H in Class 0 objects are found for H$_2$O, being significantly lower than those for methanol and the rest of species. 

In Fig. \ref{Fig: Double deuteration}, we compare the results of the double to single deuterated ratios of CH$_3$OH, H$_2$O, and H$_2$S. 
In the case of these species that form on the grain surfaces, the relation D$_2$/D = 0.25 D/H would be hold as long as the distribution of D atoms is purely statistical, meaning just proportional to the D/H ratio at formation \citep{Ceccarelli2014}.
Our first result is that the D$_2$/D ratio is higher than the statistical value for all of them. Moreover, the deuteration is very different from one compound to another, CH$_3$OH being the one with higher deuteration levels, followed by H$_2$S, and H$_2$O, which is the one with the lowest deuteration values.
The difference between the deuteration of H$_2$O and CH$_3$OH has been interpreted by different authors in terms of an evolutionary sequence  (see, e.g, the review by \citealp{Ceccarelli2014}): the O hydrogenation leading to water occurs first, and, once formed, water remains frozen on the grains \citep{Hollenbach2009}. The formation of H$_2$CO by CO hydrogenation, and subsequently the formation of CH$_3$OH, occurs later \citep{Cazaux2011,Taquet2012b}. Since the deuteration of molecules on the grain surface is governed by the gas D/H ratio at formation, which increases with decreasing temperature along the evolution of the pre-stellar core, larger deuterium fractions correspond to larger D/H gas ratios, and therefore to a later formation and deuteration of the molecules. 

This explanation may work for CH$_3$OH and H$_2$O, but it would not explain the difference between H$_2$O and H$_2$S since both are expected to form in the first stages of the cloud collapse. In order to understand the difference between these two species, we need to take into account the occurrence of H-D and D-H substitution reactions on the grain surfaces \citep{Oba2012, Oba2019}. H$_2$O is only deuterated during its formation process \citep{Oba2012,Taquet2013a}, while H$_2$S can also be deuterated through H–D substitution processes on the grain surfaces after their formation \citep{Oba2019}. The reverse reaction, D-H substitution, can also occur, thus it could be thought that this mechanism may in principle diminish or suppress the deuteration level of H$_2$S. However, laboratory experiments carried out by \citet{Oba2019} show that chemical desorption is an effective desorption mechanism of H$_2$S and its deuterated compounds. Once HDS is formed from H$_2$S via H–D substitution, it could be released to the gas phase via chemical desorption, while on the contrary the desorbed HDS undergoes D-H substitution only when readsorbed onto the grain surfaces. Therefore this mechanism may increase the deuteration levels of H$_2$S in the gas phase, potentially exceeding the atomic D/H ratio in the environment at formation. The efficiency of this mechanism in the ISM conditions should be further explored. 

Surface H-D substitution reactions may also occur in the case of methanol to form CH$_2$DOH and CHD$_2$OH. In this case, the extremely low occurrence of D–H substitution reactions \citep{Nagaoka2007} could help to maintain high levels of deuteration in these species in gas phase. In other words, once CH$_2$DOH and CHD$_2$OH are formed through the H–D substitution reactions of CH$_3$OH,  it could be difficult to return to non-deuterated methanol. In order to confirm this interpretation, it is interesting to compare the deuteration on oxygen (CH$_3$OD) with that in carbon ( CH$_2$DOH and CHD$_2$OH). Since abstraction of the H from OH is not possible, the CH$_3$OD/CH$_3$OH should be significantly lower than the CH$_2$DOH/CH$_3$OH ratio, and closer to the HDO/H$_2$O ratio. This is indeed in agreement with observational data, which shows that (D/H) for CH$_3$OD is significantly lower by more than one order of magnitude, than for CH$_2$DOH (see Fig. \ref{Fig: DH molecules and sources}).
This supports the interpretation that the differences in the deuteration of H$_2$O, H$_2$S, and CH$_3$OH are not only due to differences in the D/H ratio at their formation but  also to the subsequent H–D and D-H substitution reactions. 

Deuteration is a complex process which depends on the initial physical conditions and H$_2$ OPR, as well as on the evolution of the molecular cloud during its collapse to form protostars. To explain, at least qualitatively, the observed D$_2$/D and D/H values in H$_2$O, H$_2$S, and CH$_3$OH, the H-D and D-H substitution reactions on the grain surface and chemical desorption need to be considered. A detailed chemical model including all these effects is necessary to have a deeper insight into the evolution of deuteration during the star formation process.

\section{Summary and conclusions}
\label{Sec: Summary and conclusions}

This work is aimed at investigating the deuteration of H$_2$S in cold cores. Although HDS was first detected by \citet{vanDishoeck1995} towards the cold star-forming core IRAS 16293 almost two decades ago, the deuterium fraction of H$_2$S in starless cores had not been studied thus far. We use observations of the GEMS IRAM 30m Large Program, as well as complementary observations with the 30m IRAM telescope. We have searched for HDS and D$_2$S in 19 starless cores selected from the whole GEMS sample, located in Taurus, Perseus and Orion molecular clouds, characterised by having different star formation activities. HDS emission has been detected in 10 of these starless cores, and D$_2$S emission has been detected in 5, which significantly increases the number of detections of these compounds in star forming regions. Upper limits have been derived for the rest of the sample. The HDS and o-D$_2$S column densities have been derived using the RADEX code and the recent HDS and o-D$_2$S collisional coefficients published by \citet{Dagdigian2022}, providing the most accurate values of these compound abundances. 

We have analysed the H$_2$S single and double deuterium fraction with regard to their relation with the corresponding cloud physical parameters, their comparison with values obtained for other interstellar sources, and their comparison with observed deuterium fractions observed of other molecules which may be related to H$_2$S in terms of formation paths or the influence of sulphur abundance. Our results can be summarised as follows: 

\begin{itemize}

\item
Regarding the relation between H$_2$S deuteration and the cloud physical conditions, we find that the value of the HDS/H$_2$S abundance ratio shows a decreasing trend with increasing cloud kinetic temperature, taking values of 0.1$-$0.3 in starless cores with T $<$ 18~K. This relation is expected, since low temperatures favor higher deuteration values. We have not detected HDS in warmer cores, which show upper limits of X(HDS)/X(H$_2$S) $<$0.03. No trend of X(HDS)/X(H$_2$S) is found with molecular hydrogen density and/or visual extinction. The five starless cores with D$_2$S detections show values of X(D$_2$S)/X(HDS)$\sim$0.05$-$0.3. The low number of detections prevents us to carry out a reliable study of the dependence of this ratio with the cloud physical conditions.

\item
The comparison of the HDS/H$_2$S in this work sample with other interstellar sources from the literature reveals a similar range of values with respect to Class 0 sources. This may be the consequence of an observational bias where observed values for Class 0 sources using single-dish telescopes are still produced by their cold envelopes. Interferometric observations will be needed to clarify this issue. The comparison of D$_2$S/HDS abundance ratio with other interstellar sources suggests that this ratio is increasing along the starless core evolution, but the limited number of detections and sources prevents a firm conclusion. 

\item
We have compared the H$_2$S deuterium fraction with that of H$_2$O, H$_2$CS, H$_2$CO, c-C$_3$H$_2$, and CH$_3$OH in other interstellar sources. As a general trend, the deuteration of all these compounds decreases with temperature during the star formation process, with lower values in Class I sources and massive star forming regions. Focusing on starless cores and young Class 0 objects, H$_2$CS and H$_2$CO present higher deuterium fractions that H$_2$O, H$_2$S, c-C$_3$H$_2$, and CH$_3$OH. We interpret this difference as the consequence of different formation and deuteration processes: H$_2$CS and H$_2$CO are mainly formed and deuterated through gas-phase reactions and on grain surfaces.The cyclic hydrocarbon  c-C$_3$H$_2$ is thought to be formed in gas phase while H$_2$O, H$_2$S, and CH$_3$OH are thought to form on the grain surfaces. Interestingly enough, water single and double deuterations are one order of magnitude lower than those of H$_2$S, even though both H$_2$O and H$_2$S are formed by hydrogenation of the O and S atoms, respectively. Higher deuterium fractions are observed in the case of CH$_3$OH.

\item
We propose that the observed behaviours of the D$_2$/D and D/H values in H$_2$O, H$_2$S, and CH$_3$OH are the consequence of a complex surface chemistry involving H-D and D-H substitution, different chemical desorption efficiencies, and the chemical evolution of starless cores. While H$_2$O is only deuterated during its formation process, H$_2$S can be deuterated, afterwards, through H–D substitution processes on the grain surfaces. The reverse reaction, D-H substitution, can also occur in the grain surfaces. However, after its formation, HDS can be released into the gas phase via chemical desorption, thus keeping high deuteration levels of H$_2$S in the gas \citet{Oba2019}. The processes are similar for deuterated methanol (on carbon) but the relative importance are different from H$_2$S. In the case of methanol, once deuterated methanol such as CH$_2$DOH and CHD$_2$OH is formed through the H–D substitution reactions of CH$_3$OH, and the inverse reactions have lower rates than in the case of H$_2$S  \citet{Nagaoka2007}.  Abstraction of the H from OH is not possible, for this reason the CH$_3$OD/CH$_3$OH is  significantly lower than the CH$_2$DOH/CH$_3$OH ratio.

\end{itemize}

This work confirms once again that deuterium fraction provides essential information about the formation and evolution of interstellar molecules. Further specific interferometric observations are needed to keep clarifying the evolution of deuteration along the star formation processes, while the full understanding of the deuteration processes requires laboratory experiments that help to understand the complex ice chemistry.

\begin{acknowledgements}
We thank the Spanish Ministerio de Ciencia e Innovaci\'on for funding support through project PID2019-106235GB-I00. AF is grateful to the European Research Council (ERC) for funding under the Advanced Grant project SUL4LIFE, grant agreement No 101096293. DNA acknowledges funding support from Fundaci\'on Ram\'on Areces through its international postdoc grant program. RMD has received funding from "la Caixa" Foundation, under agreement LCF/BQ/PI22/11910030. I.J.-S. has received support from grant No. PID2019-105552RB-C41 by the Spanish Ministry of Science and Innovation/State Agency of Research MCIN/AEI/10.13039/501100011033. We thank the anonymous referee for valuable comments that improved the manuscript.
\end{acknowledgements}

\bibliography{gems_HDS}

\appendix

\end{document}